\documentclass[draft]{agujournal2019}
\usepackage{url} %this package should fix any errors with URLs in refs.
\usepackage{lineno}
\usepackage{amsmath}
\usepackage[finalnew]{trackchanges} %for better track changes. finalnew option will compile document with changes incorporated.
\usepackage{soul}

\draftfalse

\journalname{Space Weather}

\begin{document}

%% ------------------------------------------------------------------------ %%
%  Title
%
% (A title should be specific, informative, and brief. Use
% abbreviations only if they are defined in the abstract. Titles that
% start with general keywords then specific terms are optimized in
% searches)
%
%% ------------------------------------------------------------------------ %%

\title{Quantifying the uncertainty in CME kinematics derived from geometric modelling of Heliospheric Imager data}

%% ------------------------------------------------------------------------ %%
%
%  AUTHORS AND AFFILIATIONS
%
%% ------------------------------------------------------------------------ %%

% Authors are individuals who have significantly contributed to the
% research and preparation of the article. Group authors are allowed, if
% each author in the group is separately identified in an appendix.)

% List authors by first name or initial followed by last name and
% separated by commas. Use \affil{} to number affiliations, and
% \thanks{} for author notes.
% Additional author notes should be indicated with \thanks{} (for
% example, for current addresses).

% Example: \authors{A. B. Author\affil{1}\thanks{Current address, Antartica}, B. C. Author\affil{2,3}, and D. E.
% Author\affil{3,4}\thanks{Also funded by Monsanto.}}

\authors{L. Barnard\affil{1}, M. J. Owens\affil{1}, C. J. Scott\affil{1}, M. Lockwood\affil{1}, C. A. de Koning\affil{2,3}, T. Amerstorfer\affil{4}, J. Hinterreiter\affil{4}, C. M\"{o}stl\affil{4,5}, J. A. Davies\affil{6}, P. Riley \affil{7}}

\affiliation{1}{Department of Meteorology, University of Reading, Reading, UK}
\affiliation{2}{Cooperative Institute for Research in Environmental Sciences, University of Colorado Boulder, Boulder, Colorado, USA}
\affiliation{3}{Space Weather Prediction Center, NOAA, Boulder, Colorado, USA}
\affiliation{4}{Space Research Institute, Austrian Academy of Sciences, Schmiedlstrasse 6, 8042 Graz, Austria}
\affiliation{5}{Institute of Geodesy, Graz University of Technology, Steyrergasse 30, 8010 Graz, Austria}
\affiliation{6}{RAL Space, Rutherford Appleton Laboratory, Harwell Campus, Didcot, OX11 0QX UK}
\affiliation{7}{Predictive Science Inc., San Diego, CA, USA}
\correspondingauthor{Luke Barnard}{l.a.barnard@reading.ac.uk}

%% Keypoints, final entry on title page.

\begin{keypoints}
\item We test the performance of geometric models for estimating coronal mass ejection kinematics with a suite of solar wind numerical model runs.
\item For Earth-directed coronal mass ejection scenarios, geometric modelling errors are minimised for observers in the L5 region.
\item Geometric modelling generally overestimates a coronal mass ejections speed and predicts earlier arrivals at Earth by, on average, 8 hours.
\end{keypoints}

%% ------------------------------------------------------------------------ %%
%
%  ABSTRACT and PLAIN LANGUAGE SUMMARY
%
%% ------------------------------------------------------------------------ %%

\begin{abstract}
Geometric modelling of Coronal Mass Ejections (CMEs) is a widely used tool for assessing their kinematic evolution. Furthermore, techniques based on geometric modelling, such as ELEvoHI, are being developed into forecast tools for space weather prediction. These models assume that solar wind structure does not affect the evolution of the CME, which is an unquantified source of uncertainty.
We use a large number of Cone CME simulations with the HUXt solar wind model to quantify the scale of uncertainty introduced into geometric modelling and the ELEvoHI CME arrival times by solar wind structure. We produce a database of simulations, representing an average, a fast, and an extreme CME scenario, each independently propagating through 100 different ambient solar wind environments. Synthetic heliospheric imager observations of these simulations are then used with a range of geometric models to estimate the CME kinematics. The errors of geometric modelling depend on the location of the observer, but do not seem to depend on the CME scenario. In general, geometric models are biased towards predicting CME apex distances that are larger than the true value. For these CME scenarios, geometric modelling errors are minimised for an observer in the L5 region. Furthermore, geometric modelling errors increase with the level of solar wind structure in the path of the CME. The ELEvoHI arrival time errors are minimised for an observer in the L5 region, with mean absolute arrival time errors of $8.2\pm1.2$~h, $8.3\pm1.0$~h, and $5.8\pm0.9$~h for the average, fast, and extreme CME scenarios.
 \end{abstract}

\section*{Plain Language Summary}
\add[LB]{Coronal Mass Ejections (CMEs) are the largest space weather hazard to society. To help manage this hazard, we need to understand how CMEs flow through space and to develop methods to forecast when they will arrive at Earth.
To help understand how CMEs flow, a range of geometric models have been developed and are widely used. Geometric models approximate a CME as a simple geometric shape, such as a circle or ellipse, and are used to help interpret CME remote sensing observations from heliospheric imagers.
So far, it has been difficult to work out how good the assumptions of geometric models are and how uncertain their predictions are. In this study, we use numerical simulations of the solar wind and CMEs to try and estimate how good the geometrical modelling assumptions are, and the size of the uncertainties on their predictions.
We find that because the geometric models don't account for time-dependent solar wind structure, that they are biased and typically predict that a CME is further out into the solar wind than it actually is. Because of this, geometric models tend to predict early arrival times at Earth.}

\section{Introduction}
Coronal Mass Ejections (CMEs) are eruptions of magnetised plasma from the Sun's atmosphere, which then propagate \change[CADK]{out}{outward} through the heliosphere and solar wind \cite{Webb2012}. CMEs play a central role in the evolution of the Sun's magnetic field and the heliosphere \cite{Owens2013a}, and they are also the main driver of severe space weather throughout the solar system, but particularly at Earth \cite{Cannon2013,hapgood_development_2020}. Consequently the study of CMEs is important from both the space science and space weather perspectives \cite{the_space_weather_editors_space_2021}.

For example, effective space-weather forecasting requires the observation and modelling of the evolution of CMEs, to predict not only CME arrival times at Earth, but also CME properties such as arrival speed \cite{Owens2020b}. A critical part of this processes is the interpretation of remote sensing observations of CMEs, particularly from white-light coronagraph and heliospheric imager instruments, such as those aboard NASA's STEREO spacecraft\cite{Kaiser2008,Howard2008}, and more recently aboard the Parker Solar Probe and Solar Orbiter missions \cite{Vourlidas2016,howard_solar_2020}.

This is a challenging problem as the evolution of CMEs through the solar wind and heliosphere is still poorly understood, due to historically sparse heliospheric observations and many open questions regarding CME structure \cite{Luhmann2020}. \citeA{riley_sources_2021} demonstrated that uncertainties in CME arrival time predictions are limited by observational uncertainties on CME parameters such as mass, speed, and direction, as well as uncertainty on the ambient solar wind structure. Since the advent of the SMEI mission in 2003 and NASA's STEREO mission in 2006, we have been able to routinely observe the propagation of CMEs from the Sun to Earth-like distances with the white-light Heliospheric Imager (HI) instruments \cite{Eyles2008,Harrison2017,Webb2006}. However, interpreting HI observations in terms of CME position, speed, and morphology, and subsequently using these parameters for forecasting, is difficult owing to observational degeneracies; that is, different combinations of CME properties that can create similar features in HI data.

\begin{figure}
\includegraphics[width=\textwidth]{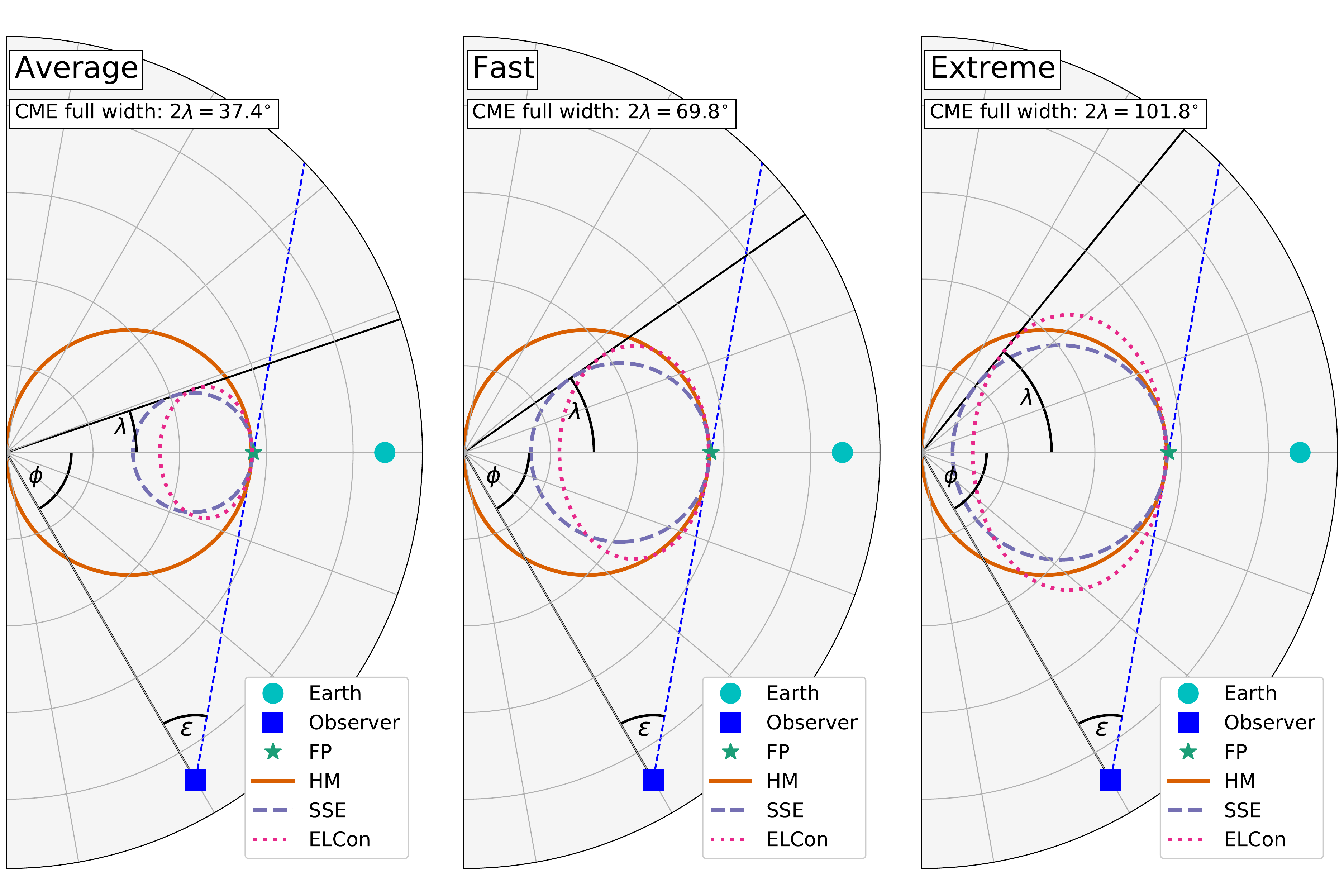}
\caption{A diagram of the different classes of geometric model used in this study, including the FP, HM, SSE, and ELCon models. The results of these models are shown for each of the considered CME scenarios, for an L5 observer, at an observed CME flank elongation of $40^{\circ}$.}
\label{fig:f01}
\end{figure}

Despite these challenges, a now mature class of model has been developed to interpret the HI CME observations in terms of CME position, speed, and morphology in a plane using simple 2D geometric shapes; these are typically referred to as CME geometric models. Examples of such models include the Point-P, Fixed-Phi (FP), Harmonic Mean (HM), Self-Similar Expansion (SSE), and ELiptical Conversion (ELCon) models \cite{kahler_v_2007,Sheeley1999,Rouillard2008,Lugaz2009,Davies2012,Mostl2013,Mostl2015,Rollett2016}. Within these models it is assumed that the maximum radial angular coordinate (elongation) of a CME feature observed in a HI image corresponds to a line of sight that is tangent to that CME's flank. This line of sight is then used to locate a geometric shape, such as a point, circle, or ellipse, that approximates the CME's spatial extent in 2D. Figure \ref{fig:f01} is a schematic that shows examples of these different geometric models, for an observer at the L5 Lagrange location recording the CME flank to be at $40^\circ$ elongation. The L5 Lagrange point is a stable gravity-well in which it is possible to maintain a stable orbit, and is the expected location for ESA's future operational space weather monitor mission, currently called Lagrange.

Hence, with a sequence of HI images, such methods allow us to compute estimates of a CME's kinematics. These kinematic profiles have been used to study the physics of CME evolution \cite{Mishra2012,Mishra2014,Harrison2012,rollett_constraining_2012}, and also as a component of models to forecast CME arrival at Earth, for example ELEvoHI \cite{Rollett2016,amerstorfer_ensemble_2018}. However, there are significant uncertainties associated with this CME geometric modelling framework, relating to both the processing of the HI observations \cite{Williams2009,Barnard2015a}, and assumptions of the models \cite{Barnard2017a}.

Tracking the CME feature in HI images is typically done manually, via time-elongation maps made along a fixed position-angle, which introduces uncertainty and subjectivity \cite{Williams2009,Barnard2015a,Barnard2017a}. Although some attempts have been made to automate the procedure, e.g. CACTUS-HI \cite{Pant2016} and J-Tracker \cite{Barnard2015a}, this is a particularly challenging problem and has only had modest success. More recently the Solar Stormwatch Citizen Science project has tried to minimise the subjectivity and quantify the uncertainty of tracking CMEs in HI data \cite{Barnard2017a} by combining many independent manual tracks. Although this work did improve the stability and reduce the subjectivity of the CME tracking, significant uncertainty still remained.

Alongside the observational challenges, at present these geometric models depend on quite severe assumptions that also introduce uncertainty into the modelling results. Most importantly, these models are ``rigid'' and do not interact with the structured solar wind \cite{hinterreiter_why_2021}. Yet it is clear from both observations and modelling that structured solar wind affects the evolution of CMEs in a non-uniform way \cite{Savani2010,Owens2017c,owens_coherence_2020}. Also, these models have between one and three free-parameters that constrain the CME direction, angular half-width, and aspect-ratio, and these are typically assumed to be stationary. Although we note it is theoretically possible to make the parameters of the geometric models time-dependent, practically it is not possible to implement this effectively due to a lack of suitable observations.

However, these models offer several advantages that make them potentially desirable for use in space-weather forecasting applications. They are computationally simple to implement and exceptionally cheap to run. Furthermore, they can provide an estimate of CME kinematics with a bare minimum of observations, requiring only coronagraph and/or HI observations from a single perspective. This is in contrast to more complex modelling procedures, such as the numerical MHD space weather forecasts produced using models such as Enlil, EUHFORIA, HelioMAS, or HelioLFM \cite{Odstrcil2003,poedts_european_2020,Riley2001,Merkin2016}; these models are much more computationally expensive and require more observational constraints to estimate the background solar wind structure, which also has significant associated uncertainty \cite{gonzi_impact_2020}.

Several studies have looked at the efficacy of using geometric models in a space-weather forecasting context. \citeA{Mostl2014} analysed a sample of 22 CMEs with the FP, SSE and HM geometric models, showing an average arrival time error of $6.1\pm 5.0$ hours, and concluded there was no evidence of one class of model performing systematically better than others. Following this, \citeA{Rollett2016} presented and assessed the ELEvoHI model, which models the CME front as an ellipse with the ELCon geometry that propagates according to the drag-based-model (DBM) \cite{Vrsnak2013}. By analysis of hindcasts of 21 CMEs, this work concluded that ELEvoHI was more skilful than the FP, SSE and HM methods, and had an average CME arrival time error of $6.4 \pm5.3$ hours. \citeA{amerstorfer_ensemble_2018} presented a new version of ELEvoHI, which includes an ensemble modelling strategy. \citeA{amerstorfer_evaluation_2021} evaluated different model set-ups of ELEvoHI by hindcasting 15 CMEs and demonstrated a mean absolute arrival time error of $6.2\pm 7.9$ hours. Similarly, \citeA{braga_predicting_2020} used a combination of the ELCon geometric model with \change[LB]{the DBM}{a parameterisation of the hydrodynamic drag on a CME} to produce hindcasts of 14 CMEs observed by HI. They found a mean absolute arrival time error of $6.9\pm 3.9$ hours. 

\citeA{mostl_modeling_2017} analysed 1337 CMEs tracked in the STEREO HI data and estimated their arrival times at different observatories using the SSE geometric model, assuming CMEs travelled with constant speed and direction. This revealed a mean arrival time error of $2.6\pm 16.6$ hours, and that for every correctly predicted ``hit'' there were two-to-three corresponding ``missess''. A review of CME forecasting techniques by \citeA{Riley2018}, many based on 3D MHD models, concluded that the mean absolute error in arrival time is around $\pm 10$ hours. In this context the average arrival time error statistics of the geometric-model-based methods appears to be comparable to, or even slightly better than, an ``average forecast''. But this would be an overly simplistic and incorrect conclusion to draw. 

For example, \citeA{Barnard2017a} produced hindcasts of 4 Earth directed CMEs observed by both STEREO-A and STEREO-B using the FP, SSE, HM and ELCon geometric models. They showed that within the observational uncertainties it was not possible to distinguish between these geometric models. Furthermore, the kinematics estimates returned from the STEREO-A and STEREO-B perspectives were very inconsistent with each other, invalidating the ``rigid'' assumption of the geometric models. In some instances, the kinematics profiles showed unphysical accelerations. Finally, the skill of these hindcasts was almost always worse than that of the operational Space Weather Prediction Center's Enlil forecast, and did not generally improve as more HI tracking data was included in the model, as would be intuitively expected if the HI data were adding forecast value.

\cite{Liu2013e} \add[LB]{used a triangulation approach with the FP and HM geometric models to assess the kinematics of 3 fast CMEs observed by STEREO-A and STEREO-B. The triangulation approach aims to reduce the uncertainty in the derived geometric modelling parameters by including the two views of a CME from the STEREO-A and STEREO-B instruments.}\cite{Liu2013e}\add[LB]{noted that although both techniques appeared to provide a fair reconstruction of the CME kinematics in the low heliosphere, both techniques systematically disagreed on the CME propagation direction, and both showed unexplained and unphysical late stage accelerations.} \cite{liu_sun--earth_2016} \add[LB]{extended this work to also consider 3 slow CMEs. They concluded that the FP and HM triangulation techniques also provide a fair representation of the kinematics slow CMEs but they also highlighted 3 important points; firstly, the results can depend sensitively on how the STEREO HI data are processed to extract the CMEs time-elongation profile; secondly, the CME propagation direction is best estimated by the FP technique in the low heliosphere, and by the HM technique at distances >100 solar radii; finally, that the FP model more often returns unphysical late stage accelerations, and in these instances it is better to use the HM technique}

Most recently, \citeA{hinterreiter_why_2021} assessed the performance of ELEvoHI for hindcasts of 12 CMEs using the STEREO-A and STEREO-B HI observations independently for each event. Similar to previous results, a mean absolute arrival time error of $7.5\pm 9.5$ was found. But this study also demonstrated inconsistent arrival time forecasts for CMEs forecast from the different STEREO-A and STEREO-B vantage points. This inconsistency implies a breakdown of the rigid CME structure within the geometric model.

\citeA{Lugaz2009} used synthetic observations of an MHD simulation of two-interacting CMEs as a test of the Point-P, FP and HM geometric models. They concluded these methods were valid in the low, inner heliosphere, but that errors grew significantly at larger distances. In a follow-up study of a simulated fast and wide CME, \citeA{Lugaz2011} showed that the FP and HM geometric models provided a better estimate of the CME speed than they did CME direction, also showing that the estimated CME direction is biased by the observers location relative to the CME.

Therefore, on balance, it is difficult to draw firm conclusions on the performance of geometric models for estimating CME kinematics and arrival-time forecasts. Although they can return favourable CME arrival time errors, there is clear evidence that the assumptions of the models are routinely broken. They can also return plausible arrival-time estimates which are derived from the integration of physically implausible kinematics profiles, which is a scientific quagmire. Additionally, the convolution of observational and model uncertainties, confounded by only modest sample sizes in the discussed statistical studies, means that not only is it unclear precisely how well the geometric models perform, it is also unclear why they perform as they do. \add[LB]{This issue was highlighted by }\citeA{howard_application_2010}, \add[LB]{who discussed how the challenge of improving CME geometric modelling depends on two coupled problems; the physics describing the appearance of a CME in observations; and the physics governing a CME's evolution. Future progress in improving CME geometric models necessarily involves advancing on each of these problems jointly.}

Here we aim to test the efficacy of geometric modelling using synthetic observations of simulated CMEs. By using simulations of CMEs evolving through realistic time-dependent and structured solar wind, observed simultaneously from a range of heliospheric longitudes, we will provide a robust quantification of the uncertainty in geometric modelling due to solar wind structure and observer location, absent of observational uncertainty. To do this, we construct three CME scenarios, representing an average, a fast, and an extreme CME. The evolution of these CMEs are modelled as Cone CMEs with the Heliospheric Upwind Extrapolation with time dependence (HUXt) solar wind model \cite{Owens2020}, through 100 ambient solar wind solutions, where the Cone CME parameterisation models CMEs as a purely hydrodynamic perturbation. With these simulation data we analyse the errors in estimating the CME apex kinematics with the FP, HM, SSE, and ELCon geometric models constrained with observations from a single observer. We also evaluate the performance of the ensemble ELEvoHI CME forecasting system \cite{amerstorfer_evaluation_2021} for these simulated CME scenarios. We focus on single-spacecraft geometric modelling techniques, rather than any stereoscopic techniques, as it seems likely that there will be at most one operationally focused heliospheric imager for future space weather forecasting, through, for example, ESA's Lagrange mission currently under development \cite{kraft_remote_2017, gibney_space-weather_2017}.

The formulation of the CME scenarios is described in Section \ref{sec:scenarios}. Section \ref{sec:models} introduces the models used in this experiment; the geometric models in Section \ref{sec:cme_geomods}, ELEvoHI in Section \ref{sec:ELEvoHI}, and HUXt in Section \ref{sec:huxt}. \add[LB]{Our experimental design is described in Section} \ref{sec:expt_design}. The results are presented in section \ref{sec:results}, whilst our conclusions are discussed in Section \ref{sec:conclusions}.

\section{CME Scenarios}
\label{sec:scenarios}
We wish to test the performance of the geometric models for a range of CME scenarios. To do this, we construct three CME scenarios to represent an average, a fast, and an extreme CME. We derive the parameters of these scenarios from the statistics of CME parameters provided in the KINCAT catalogue of the HELCATS project \url{https://www.helcats-fp7.eu/catalogues/wp3_kincat.html}. More details of the HELCATS analysis and work-packages are provided in \citeA{barnes_cmes_2019} and \citeA{pluta_combined_2019}. The KINCAT data includes graduated cylindrical shell (GCS) fits \cite{thernisien_implementation_2011} of 122 CMEs observed in the COR2 coronagraphs.
\add[LB]{These GCS fits return estimates of the CME apex speed and the angular half-width.}

\remove[LB]{For our purposes we require the CME speed and width for each scenario; a CME will always be Earth directed in our scenarios and so}\add[LB]{In our CME scenarios, the initial CME apex will have Earth's longitude and latitude and is directed radially, and the CME has a $0^{\circ}$ inclination to the ecliptic plane. Therefore, for our purposes, we require only the CME speed and full-angular width for each scenario.} We compute \add[LB]{the} speed and \add[LB]{full} angular width values for the average, fast, and extreme scenarios by calculating the median, 85th percentile, and 95th percentiles of the \change[LB]{KINCAT}{distributions of the GCS} speeds and \add[LB]{double the GCS half-}widths \add[LB]{provided by KINCAT}. The resulting values are given in Table \ref{tab:scenarios}. \add[LB]{Here we focus on only Earth-directed CMEs, rather than CMEs directed off of the Sun-Earth line. This is because, in our judgement, the CME speed and width are more important parameters for determining the evolution of CMEs in the solar wind and their representation with geometric models. However, CMEs that are directed away from the Sun-Earth line are also a relevant concern and a future study will examine this.}

\begin{table}
\caption{CME scenarios}
\centering
\begin{tabular}{l | c c c }
 & Average & Fast & Extreme\\
\hline
Speed ($\mathrm{km~s^{-1}}$) & 495 & 1070 & 1427 \\
Full Width (deg) &  37.4 & 69.8 &  101.8 \\
\end{tabular}
\label{tab:scenarios}
\end{table}

\section{Models}
\label{sec:models}

\subsection{Single spacecraft CME geometric models}
\label{sec:cme_geomods}

CME geometric models are a set of techniques to interpret the spatial evolution of CMEs in terms of time-elongation profiles derived from coronagraph and/or heliospheric imager observations. Each geometric model interprets the CME structure as a simple, regular geometric shape, such as a point, circle, or ellipse. Here we use the Fixed Phi (FP), Harmonic Mean (HM), Self Similar Expansion (SSE) and ELliptic Conversion (ELCon) models, described below.

\subsubsection{Fixed Phi}

The FP model assumes that the feature being tracked is a point source, with no cross-sectional extent \cite{Sheeley1999,Rouillard2008}. With the elongation ($\epsilon$) of the observed feature defined as the Sun-Observer-Feature angle, and the Observer-Sun-CME apex angle defined as $\phi$, the FP model computes the radial distance of the feature as,

\begin{equation}
    r = \frac{r_{\rm{obs}}sin(\epsilon)}{sin(\phi + \epsilon)},
\end{equation}

where $r_{\rm{obs}}$ is the heliocentric distance of the observer. Figure \ref{fig:f01} shows examples of this observing geometry for the CME scenarios considered in this work, in which radial distance of the FP model is shown with a turquoise-star.

\subsubsection{Harmonic Mean}

The HM model, introduced by \citeA{Lugaz2009}, assumes a CME with circular cross-section that expands with one point tied to Sun-centre. Under these assumptions, the radial distance of the CME apex is computed as,

\begin{equation}
    r = \frac{2 r_{\rm{obs}} sin(\epsilon)}{1 + sin(\phi + \epsilon)}.
\end{equation}

Examples of the HM geometry are shown as the solid-orange lines in Figure \ref{fig:f01}.

\subsubsection{Self Similar Expansion}

The SSE model was introduced as a generalisation of the FP and HM geometries by including the angular half-width ($\lambda$) of the CME as an additional free parameter \cite{Davies2012}. Setting $\lambda$ to zero replicates the FP geometry, whilst setting it to $90^{\circ}$ replicates the HM geometry. Intermediate values of $\lambda$ describe the CME cross section as a self-similarly expanding circle of constant half-width. With the SSE model, the radial distance of the CME apex is computed as, 

\begin{equation}
    r = \frac{r_{obs}sin(\epsilon)(1 + sin(\lambda))}{sin(\lambda) + sin(\phi + \epsilon)}
\end{equation}

In Figure \ref{fig:f01}, the purple dashed lines show examples of the SSE geometry for the different CME scenarios. It is normally necessary to either assume a value for $\lambda$, or to estimate it from observations. In this work, $\lambda$ is set to be equal to half the angular width of the Cone CME scenario used in HUXt. This assumption is consistent with observations that a CME's angular width does not significantly change as it propagates through the inner heliosphere \cite{st_cyr_properties_2000,schwenn_association_2005}. 

\subsubsection{ELliptic Conversion}

The ELCon model is a further generalisation of the SSE model, in which the CME is modelled as a self-similarly expanding ellipse with constant angular half-width and aspect ratio \cite{Mostl2015,Rollett2016}. The ellipse aspect ratio is the ratio of the semi-major and semi-minor axis lengths, but following \citeA{Mostl2015} we work with the inverse aspect-ratio ($f$). Computation of the radial distance of the CME apex is more complex than for the FP, HM, and SSE models, although it does only have the one extra free parameter, $f$. However, by defining

\begin{align}
    \omega &= \pi - \epsilon - \beta,\\
    \beta &= arctan(f^{2}tan(\omega)),\\
    \theta &= arctan(f^{2}tan(\lambda)),\\
    \psi &= \frac{\pi}{2} + \theta - \lambda,\\
    \zeta &= \frac{\pi}{2} + \beta - \omega,
\end{align}

and 

\begin{align}
    \Omega_{\beta} &= \sqrt{f^{2}cos^{2}(\beta) + sin^{2}(\beta)},\\
    \Omega_{\theta} &= \sqrt{f^{2}cos^{2}(\theta) + sin^{2}(\theta)},
\end{align}

it can be shown that the radial distance of the CME apex is given by 

\begin{equation}
    r = \frac{r_{\rm{obs}}sin(\epsilon)sin(\lambda)\Omega_{\theta}\Omega_{\beta}}{sin(\psi)sin(\omega)\Omega_{\theta} + sin(\zeta)sin(\lambda)\Omega_{\beta}}\left(1 + \frac{\Omega_{\theta}sin(\psi)}{sin(\lambda)} \right).
\end{equation}

A full derivation is provided in \citeA{Rollett2016}. A value for $f$ must be assumed, and here we set $f=0.7$, representing an ellipse with a fairly flat front. This assumed value of $f$ is supported by the case study of \citeA{Mostl2015}, who estimated $f$ to be $0.71$ for a fast CME. Similarly, \citeA{janvier_comparing_2015} analysed the aspect ratio of an elliptical model fitted to magnetic clouds and and CME shocks, and found that $f\approx0.75$. Examples of the ELCon geometry are shown by the pink dotted lines in Figure \ref{fig:f01}.

\subsection{ELEvoHI}
\label{sec:ELEvoHI}
ELEvoHI is a CME forecasting system that combines the ELCon geometric model, with the drag-based-model (DBM) CME propagation tool \cite{Rollett2016, amerstorfer_ensemble_2018}. ELEvoHI takes as inputs an observed time-elongation profile of the CME flank from a heliospheric imager, as well as an estimate of the background solar wind speed, and returns a prediction of the time-evolution of an elliptical CME front. This can be used to forecast a CME arrival throughout the heliosphere. The most recent version of ELEvoHI employs an ensemble modelling strategy to provide estimates of the forecast uncertainty, varying the combinations of ELCon parameters within a range that is consistent with the HI observations.

Here, the ensemble version of ELEvoHI is used to provide predictions of when the simulated CMEs will arrive at Earth, taking as input the time-elongation profile of the CME flank as seen from the synthetic observers. The number of ensemble members in each CME scenario and background wind combination is variable, and depends on how well the ELEvoHI solution converged. As synthetic time-elongation profiles have no added observational noise, and are at much higher cadence than actual HI observations (~4~min timestep in the simulations, 40~minutes cadence for STEREO-HI1 images, and 120~minutes cadence for STEREO-HI2 images), the synthetic observations were degraded to a level chosen to work well with the ELEvoHI fitting procedure.
In order to estimate the drag-parameter from HI observations, the ELCon kinematics of each CME are fitted between a distance of $\sim45$ and $\sim120$ R$_\odot$ using a drag-based equation of motion.
As input for the ambient solar wind speed we use 19 different values between 250 and 700 km~s$^{-1}$ for each of which a fit is performed. The best combination of the resulting drag-parameter and solar wind speed, i.e.\ the fit with the smallest mean residual, is then used for the prediction. With the best fit solar wind speed and drag parameter, the DBM is then used to extrapolate the ELCon kinematics to provide a prediction of the CME propagation through the heliosphere. For more information on that approach see the detailed description in \citeA{amerstorfer_evaluation_2021}.

\subsection{HUXt}
\label{sec:huxt}
HUXt \cite{Owens2020} is a numerical model of the solar wind that uses a reduced physics approach, treating the solar wind as a 1D incompressible hydrodynamic flow. This allows very efficient computational solutions, being approximately 1000 times faster than comparable 3D MHD solar wind models. Despite this reduced physics approach, HUXt has been shown to closely emulate full 3D MHD models; a 40-year validation test of ambient solar wind from HelioMas \cite{Riley2001} was reproduced to within $7\%$ of the HelioMAS solar wind speeds throughout the entire model domain \cite{Riley2011,Owens2020}. Therefore, HUXt can serve as an effective surrogate in situations where full 3D MHD simulations are too computationally expensive. HUXt only requires the solar wind speed at the model inner boundary to be specified and so can work with the output of any coronal model that can provide this. Here we use output from the MAS coronal model \cite{Riley2001}, but it can also operate with output from, for example, the Wang-Sheely-Arge model \cite{Arge2000a}. In this work, HUXt is run in the latitudinal plane corresponding to Earth's latitude, in Heliospheric-Earth-Equatorial coordinates, at the initialisation time of each cone CME.

\add[LB]{In this work, the domain of HUXt is configured as follows; the radial grid has inner and outer boundaries of $30~R_{\odot}$ and $240~R_{\odot}$, with a grid step of $1R_{\odot}$; the longitude grid spans $\pm90^{\circ}$, with a grid step of $0.7^{\circ}$. The model time step is set by the Courant-Friedrichs-Lewy condition and is $232$~seconds, corresponding to a maximum speed of $3000~kms^{-1}$.}

Within HUXt, CMEs are parameterised as cone CME perturbations to the solar wind speed. Spatially, the cone CME consists of two hemispheres connected by a cylinder; at one extreme a CME is initially spherical, while more generally it is sausage shaped. The axis of the cone CME is directed radially, and located by the CME's source longitude and latitude. The CME's width is used to parameterise the angular extent of the hemispheres, while the ``thickness'' sets the length of the cylindrical portion that connects the hemispheres. This structure is advected through the model inner boundary at the CME's speed, and anywhere on the boundary within the cone CME domain is assigned the CME speed. This perturbation then propagates hydrodynamically through the model solution. This same approach is used in many 3D MHD forecasts of CMEs.

CMEs are tracked through the HUXt solution by inserting test particles into the flow on the CME surface at the model inner boundary. These test particles then passively advect with the flow and are followed at all time steps out to the model's outer boundary. In this work we must also calculate the \add[LB]{time-elongation profile of the} flank of the CME from an observer's perspective. This is done by computing the elongation of each particle on the CME boundary and finding the particle with maximum elongation in an observer's field of view. \add[LB]{Ideally, the time-elongation profile of the flank would be computed from Thomson scattering simulations of the HUXt output, forward modelling what we could expect to see from an instrument such as STEREO-HI. However, as HUXt is derived from incompressible hydrodynamics, it does not solve for the plasma density, which is a necessary requirement for the forward modelling comparison. Therefore, tracking the maximum elongation of the CME tracer particles is a necessary approximation. However, we note that} \cite{barnard_ensemble_2020} \add[LB]{showed that this approach returned time-elongation profiles that compared favourably to those extracted directly from STEREO-HI images, which gives us some confidence this approximation is reasonable.}

\add[LB]{The speed of the simulated cone CME apex is computed by calculating the numerical gradient of the time distance profile of the test particle injected on the CME apex. Similarly, the speeds computed for each geometric model perform the same numerical gradient calculations on the time distance profile of the geometrically modelled CME apex. The numerical gradient scheme uses 2nd order central differences for the interior points, and 1st order forward and backward differences at each edge. To avoid the possibility of spurious artefacts at the edges of each gradient calculation, we ignore the first and last point in each speed profile.}

\section{Experiment Design}
\label{sec:expt_design}
Our experiment is designed to assess the performance of geometric models in reconstructing the kinematics of CMEs flowing through structured solar wind. To do this, we produce a database of HUXt runs for our average, fast, and extreme cone CME scenarios, for a range of background solar wind environments. Specifically, we select 100 random model initialisation times between 2008-01-01 and 2016-01-01, sampling uniformly across this period. The MAS solar wind speed is used for each selected time. A HUXt run is then produced for each CME scenario and initialisation time \add[LB]{, with a CME being initialised at the inner boundary 1 hour after the model initialisation}. For each run, synthetic observers track the elongation of the CME flank from their perspective. The observers are at the same heliocentric distance and latitude as Earth, but at fixed longitude separations from Earth, with one observer at each $10^{\circ}$ of longitude, from $10^{\circ}$ to $90^{\circ}$ behind Earth. The L5 point corresponds to $60^{\circ}$ behind Earth, and we also compute the observations from the L4 point, $60^{\circ}$ ahead of Earth.

With each simulated time-elongation profile, we use the single-spacecraft geometric modelling techniques to estimate the kinematics of the CME apex. For each geometric model, the CME direction, $\phi$, and angular half-width, $\lambda$, are fixed and known from the CME scenarios, and are the same as those used in the Cone CME parameters. The only assumed parameter is the inverse ellipse aspect ratio ($f$) for the ELCon model, which is $f=0.7$, although we note this assumption is supported by some observational studies \cite{Mostl2015,janvier_comparing_2015}. Then the CME kinematics estimates returned by each geometric model and observer location can be compared with the true CME kinematics derived from the HUXt simulation.

We note here a potential limitation to our methodology, concerning how representative Cone CMEs in HUXt are of real CME solar-wind interactions. Because the Cone CMEs are purely hydrodynamic perturbations, we think it is plausible that they overestimate the impact of solar wind structure on CME evolution. This is because we expect the magnetic structure of CMEs would tend to resist deformation by solar wind structures. Consequently, we expect that these experiments probably represent an upper bound to the impact of solar wind structure on CMEs.

\section{Results}
\label{sec:results}
Here we detail the results of our simulation work. Before proceeding to an analysis of the whole database of runs, we first analyse some examples that highlight some of the potential issues with geometric modelling.

\subsection{Example 1: A uniform solar wind background}
\label{sec:example1}

As a first test of how well geometric models can reproduce the true CME kinematics, we explore the most simple example of the Cone CME scenarios propagating through a uniform solar wind background, where the inner boundary conditions are $400~\mathrm{km~s}^{-1}$ everywhere. This configuration is most consistent with the intrinsic assumptions of the geometric models and serves as an example of how well they can perform when their assumptions are met.

\begin{figure}
\includegraphics[width=\textwidth]{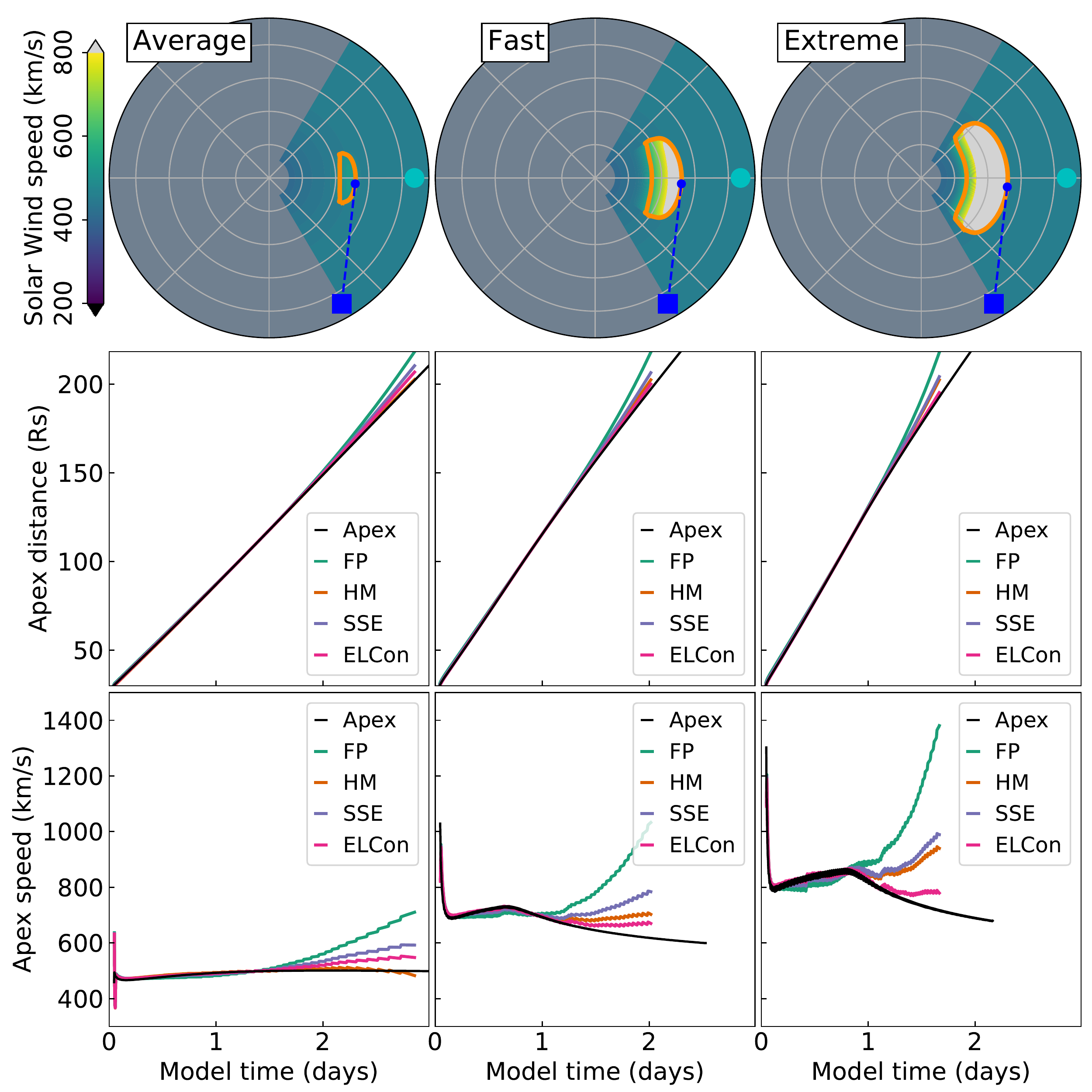}
\caption{Kinematics profiles of CME scenarios propagating through a uniform background solar wind in HUXt. The top row shows snapshots of the HUXt solution when the front of each CME scenario reaches $0.5$~AU. The middle row shows the time series of the radial position of the CME's apex computed directly from HUXt, and also as estimated by the suite of geometric models using the observations of the CME flank from the L5 location. The bottom row shows the time series of the velocity profile of each CME's radial apex position.}
\label{fig:f02}
\end{figure}

Figure \ref{fig:f02} presents snapshots of the HUXt solutions for these scenarios, as well as the true CME kinematics derived from the HUXt solution, and the kinematics estimates derived from the geometric models with a simulated observer at L5. The columns correspond to the average, fast, and extreme CME scenarios. The top row shows snapshots of the HUXt solution when the CME apex reach $0.5$~AU, whilst the middle row shows the time series of the CME apex distance, and the bottom row shows the time series of CME apex speed.

For this configuration it is clear that the geometric models all do a good job at reproducing the true CME kinematics over most of the model domain. In fact, the solutions appear to be essentially degenerate with each other out to distances of approximately $150~R_{s}$; past this distance, systematic differences between the geometric models arise and they all diverge from the true kinematics, except for the HM representation of the average scenario.

\subsection{Example 2: A structured solar wind background}
\label{sec:example2}

\begin{figure}
\includegraphics[width=\textwidth]{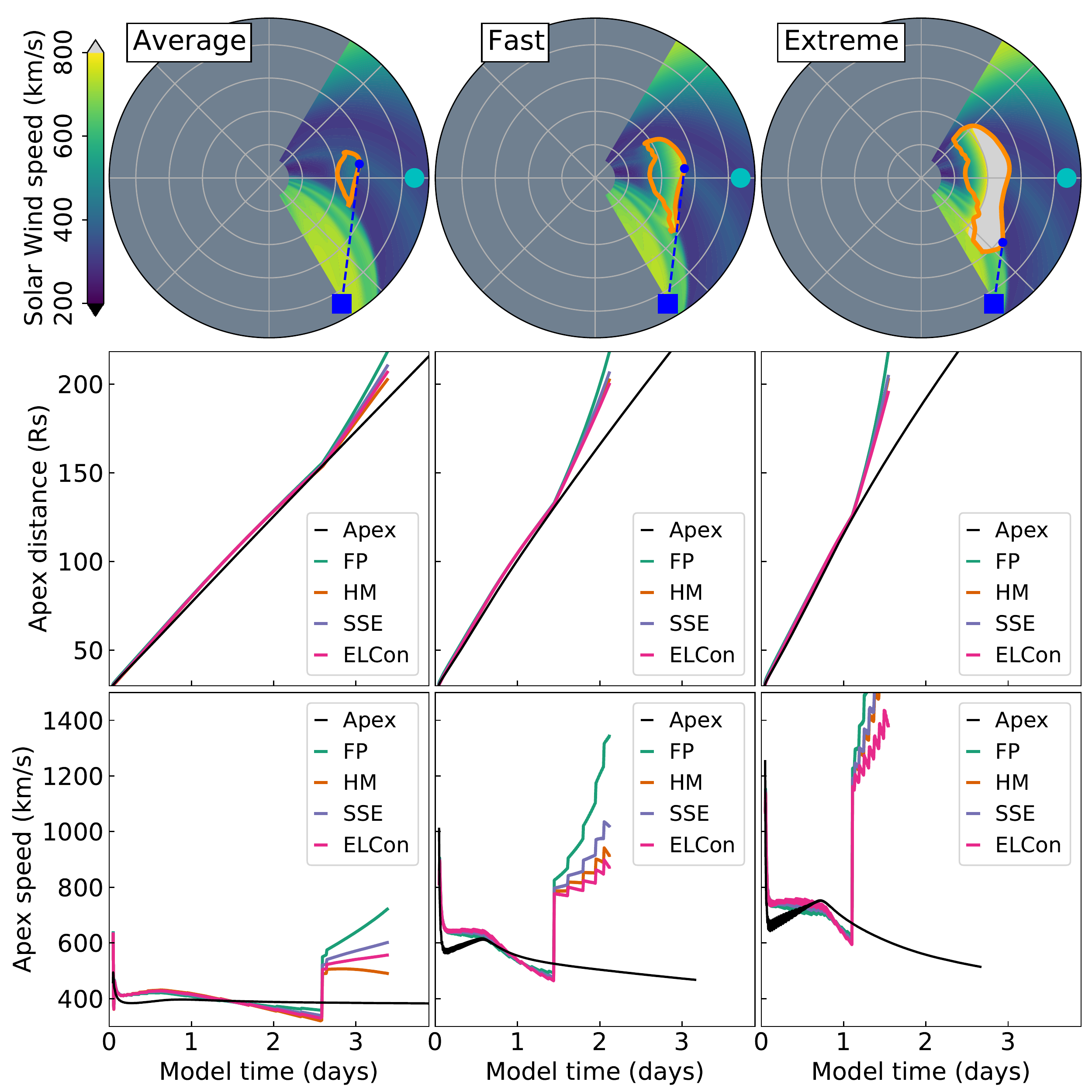}
\caption{Kinematic profiles of each CME scenario propagating through a structured background solar wind in HUXt. The background solar wind boundary condition is the HelioMAS solution for Carrington rotation 2071. The top row shows snapshots of the HUXt solution when the front of each CME scenario reaches 0.5~AU. The middle row shows the time series of the CME's radial apex position along the Sun-Earth line computed directly from HUXt, and also as estimated by the suite of geometric models using the observations of the CME flank from the L5 location. The bottom row shows the time series of the velocity profile of each CME's radial apex position.}
\label{fig:f03}
\end{figure}

In this example, instead of using a uniform solar wind speed on the HUXt inner boundary, we use the HelioMAS solution for Carrington rotation 2071, and observe the CMEs from the L5 location. Figure \ref{fig:f03} shows the simulation results for this example. In this circumstance the CME erupts into predominantly slow wind, with a fast stream just behind the eastern flank. The narrower CME in the average scenario doesn't interact strongly with this fast stream, whereas the wider CMEs in the fast and extreme scenarios do interact with the fast stream.

For each scenario, all the geometric models return essentially degenerate kinematics estimates out to approximately $150~R_{s}$. But, although the geometric models agree with each other, they are systematically different from the CME's true kinematics, which is seen most clearly with the time-speed profiles. Furthermore, in each scenario there is a discontinuity in the kinematics profiles, which occurs at around 2.5, 1.5, and 1.1 days for the average, fast, and extreme scenarios, respectively. These discontinuities occur because of how the CMEs are tracked by the L5 observer. At earlier times, due to the inclination of the CME fronts, the observer tracks the western flanks, as seen in the HUXt snapshots in Figure \ref{fig:f03} for the average and fast scenarios. However, eventually the non-uniform evolution of the CME means that the observer instead tracks the eastern flank, as is seen in the HUXt snapshot of the extreme scenario. As the tracked flank point ``jumps'' from the western to the eastern flank, without any significant change to the CME speed, the observer sees a rapid increase in the rate of change of elongation. This manifests itself in the kinematics profiles as a rapid acceleration of the CME. But the acceleration is unphysical and is in fact just a symptom of a breakdown in the assumptions of these geometric models.

\add[LB]{We note that it might be challenging to resolve these discontinuities in the time-elongation profiles derived from real HI data. These simulated time-elongation profiles have a high temporal resolution of 3.8~minutes, and there are no errors introduced into the time-elongation profile from extracting it from the HI images directly, or from a J-map. This makes these discontinuities easy to discern. However, the STEREO HI1 images have a cadence of $\approx40~$ minutes, and uncertainty is introduced into the derived time-elongation profiles by averaging the HI data into a J-map, and by the manual extraction of the time-elongation profiles}\cite{Williams2009,Barnard2015a,Barnard2017a}. \add[LB]{The reduced cadence and increased uncertainty will serve to mask these rapid changes in the kinematics. Furthermore, the extended exposure time of the HI images, of around 20 minutes, introduces motion blur into the CME fronts observed in HI images, which will also smooth the observed kinematics. We also note that the line of sight integration through the CME front is also likely to make any such discontinuity less likely to occur in reality than in our model. Another consideration is the role of an observer's bias in manually tracking profiles. It is possible that an observer would naturally try to track a feature that evolves smoothly over time, based on the assumption that the leading feature corresponds to smoothly varying intersections of the CME feature. An investigation into the current methods of manual tracking of features in HI data and the biases and uncertainties this introduces would be helpful in better understanding this issue. With the next generation heliospheric imagers, e.g. PUNCH} \cite{deforest_utility_2016} \add[LB]{, and as our skill at extracting CME features in these data improve, it is possible such features could be more easily observed.}

\subsection{Example 3: A structured solar wind background with multiple observers}
\label{sec:example3}
\begin{figure}
\includegraphics[width=\textwidth]{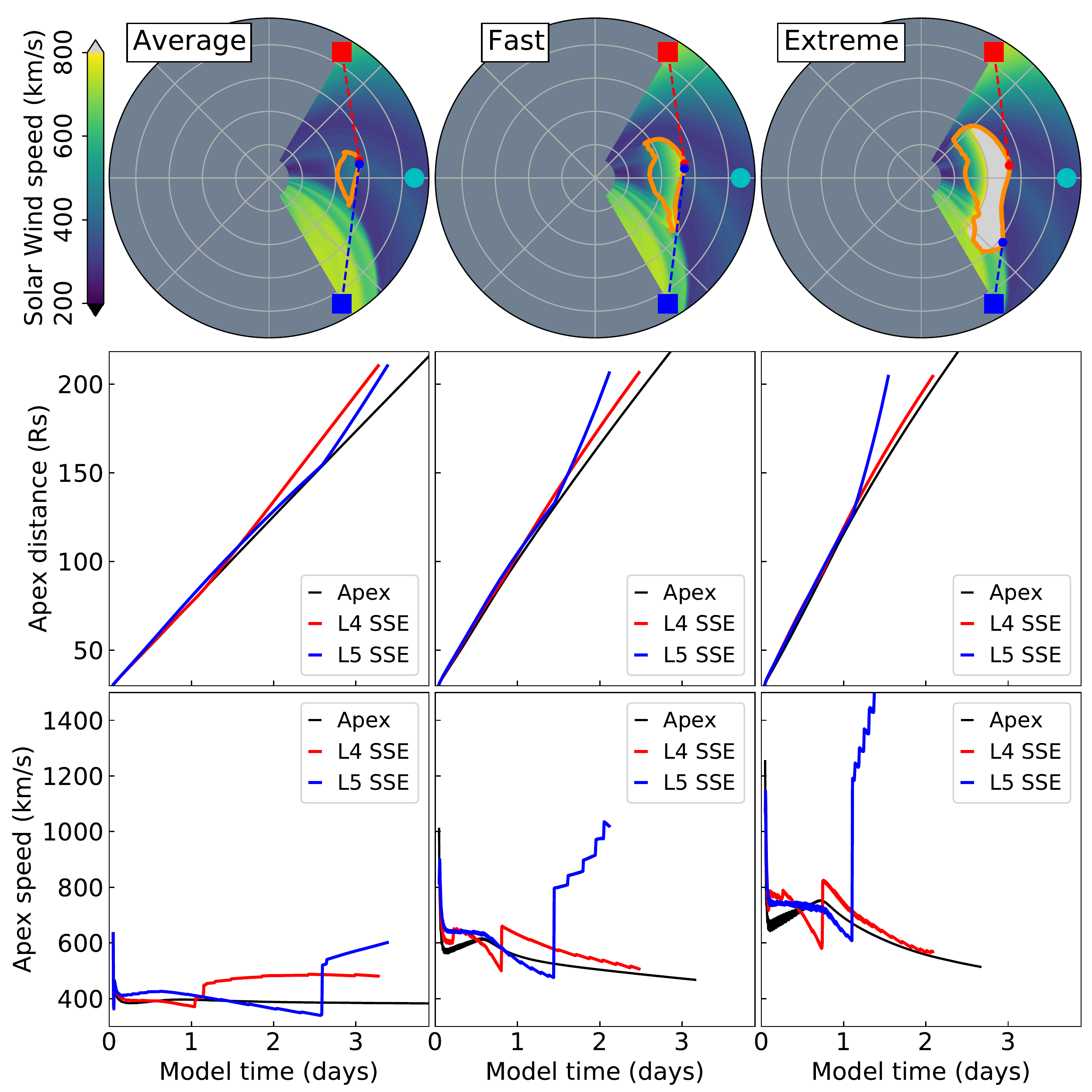}
\caption{Similar to Figure \ref{fig:f02}, these panels show the kinematics profiles of each CME scenario propagating through the structured background solar wind in HUXt, corresponding to Carrington rotation 2071. The top row shows snapshots of the HUXt solution when the front of each CME scenario reaches 0.5~AU, along with observers at the L4 (red) and L5 (blue) locations. The middle row shows the time series of the CME's radial apex position along the Sun-Earth line computed directly from HUXt, and also as estimated by the SSE geometric model using the observations of the CME flank from the L4 and L5 locations. The bottom row shows the time series of the velocity profile of each CME's radial apex position.}
\label{fig:f04}
\end{figure}

Here we repeat the same experiment as in section \ref{sec:example2}, except that we now observe the CME from both L5 and L4. These results are shown in Figure~\ref{fig:f04}. We only show the results for the SSE model for the sake of brevity, as they are qualitatively similar for each geometric model in this viewing geometry. 

For each scenario, the L5 and L4 observers return significantly different estimates of the CME kinematics. Furthermore, each observer returns systematically incorrect estimates of the CME kinematics, being biased to larger heliocentric distances than the true apex position and with different acceleration profiles. The L4 and L5 observations also show discontinuities in the kinematics profiles. These occur for the same reason as in section \ref{sec:example2}; at the discontinuity, the observed flank point jumps a large distance due to the irregular shape of the CME front. This example highlights another challenge with the geometric modelling approaches - the derived kinematics depend on the relative location of the observer to the CME, and the spatial structure on the CME front.

\subsection{Statistical analysis of all runs}
\label{sec:stats_all_runs}

Although sections \ref{sec:example1}-\ref{sec:example3} showed some illustrative examples of how geometric models can perform in specific circumstances, they are not instructive for determining the performance of geometric models in a wide range of circumstances and in an average sense. Here we will statistically analyse the geometric modelling results over all CME scenarios and background wind solutions to assess the ability of geometric models to reconstruct a CMEs kinematics more generally. 

\begin{figure}
\includegraphics[width=\textwidth]{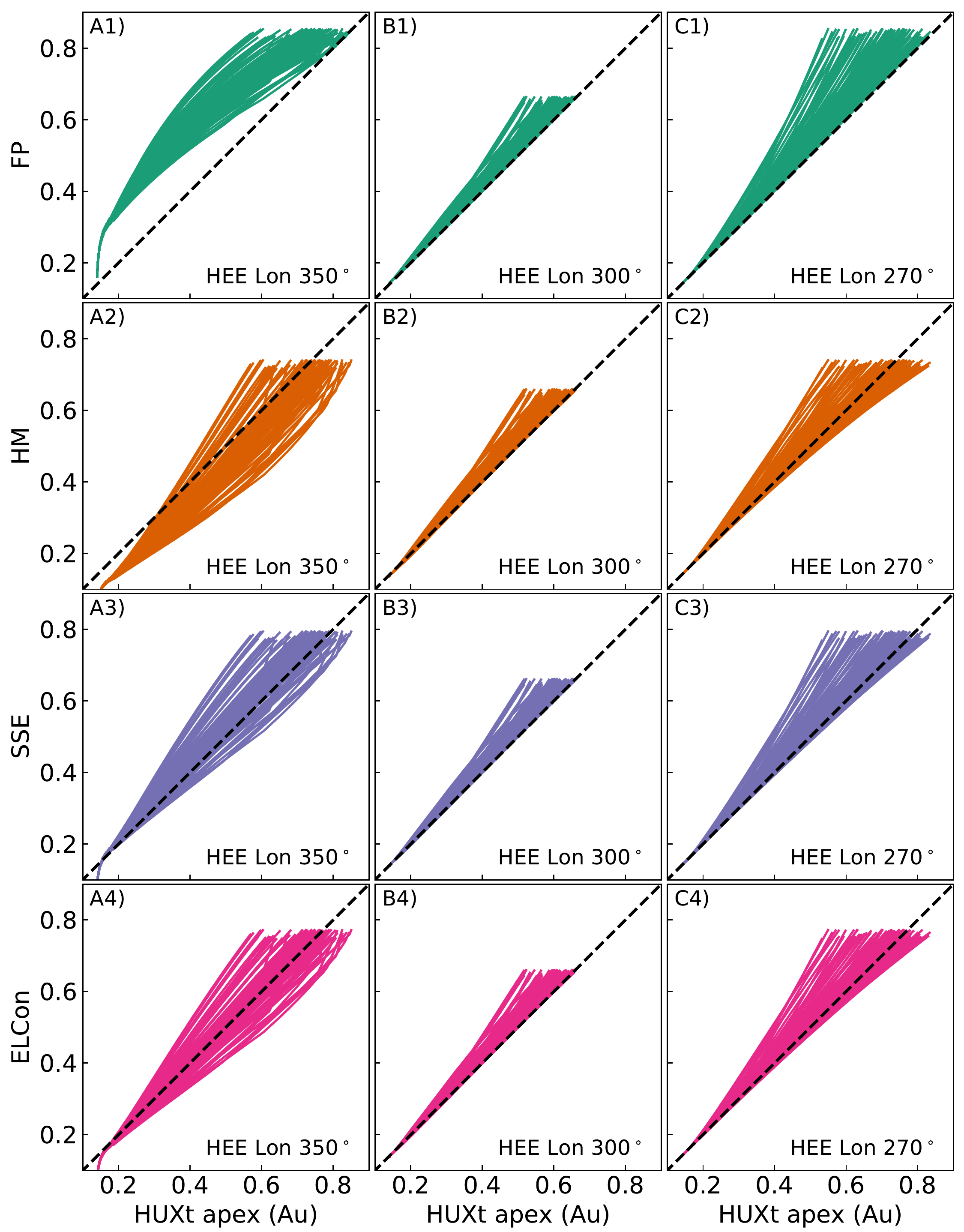}
\caption{These panels present a subset of the modelling data produced by our analysis of the three CME scenarios propagating through the 100 different ambient solar wind solutions, with each geometric model and range of observer locations. Each panel shows a comparison of the radial apex positions computed by HUXt and estimated by a geometric model, for a specific observer location. Columns A-C correspond to an observer located at HEE longitudes of $350^{\circ}$, $300^{\circ}$, and $270^{\circ}$, while each row corresponds to a different geometric model. In each panel the black-dashed line marks the one-to-one line, indicating agreement between the true and geometrically modelled CME apex position.}
\label{fig:f05}
\end{figure}

An example of the suite of modelling results that we analyse is presented in Figure \ref{fig:f05}. This plot compares the true CME apex distance with those estimated by the geometric models, for a range of different observer longitudes, for the average CME scenario only. Each row corresponds to a different geometric model, while the three columns correspond to observer longitudes of $350^{\circ}$, $300^{\circ}$, and $270^{\circ}$, respectively; these longitudes correspond to observer-Sun-CME angles ($\phi$) of $10^{\circ}$, $60^{\circ}$ (L5), and $90^{\circ}$ for these Earth-directed CME scenarios. The results for each of the 100 background solar wind solutions are shown by the colored lines in each panel. The black-dashed line is the one-to-one line, highlighting the region where the true and geometrically modelled solutions are in close agreement.

From these panels alone there are already suggestions of some conclusions regarding the performance of the geometric models. Firstly, there is a clear bias that, in general, the geometric model estimates of the CME apex radius are larger than the true value - this is illustrated by the majority of the profiles being above the one-to-one line, except for the HM model at the observer longitude of $350^{\circ}$. Secondly, the spread in the profiles is much smaller for the observer at $300^{\circ}$, than for $350^{\circ}$ or $270^{\circ}$, suggesting that the average discrepancy between the geometrically modelled kinematics and true kinematics varies as a function of the Observer-Sun-CME angle. \add[LB]{Finally, the average discrepancy between the geometrically modelled kinematics and true kinematics tends to increase as a function of the true apex distance.}

To explore this further, we will reduce these full kinematics profiles to more convenient summary statistics for each combination of CME scenario, background solar wind, geometric model, and observer location. Firstly, we compute the error of the geometrically modelled CME apex distance as,

\begin{equation}
e_{gm} = r_{gm} - r_{huxt},
\end{equation}

where $r_{gm}$ is the time-series of the CME radial apex coordinates from a geometric model, and $r_{huxt}$ is the time-series of the true CME radial apex coordinates. We then integrate $e_{gm}$ as a function of the true CME radial apex coordinate, to an outer bound of 0.5~AU, such that 

\begin{equation}
E_{gm} = \int_{0}^{0.5} e_{gm} \,dr
\end{equation}
and
\begin{equation}
H_{gm} = \int_{0}^{0.5} |e_{gm}| \,dr,
\end{equation}

where $E_{gm}$ is the integrated error, and $H_{gm}$ is the integrated absolute error. It is necessary to select an upper bound to the integration that is common to the profiles of all geometric model types and observer locations, otherwise the integration would not allow a fair comparison between them. These integrations are calculated numerically, using the trapezium rule. \add[LB]{The units of $E_{gm}$ and $H_{gm}$ are in $Au^{2}$. In this experiment we are interested in the relative variations in $E_{gm}$ and $H_{gm}$ as a function of each geometric model and observer location.}

\begin{figure}
\includegraphics[width=\textwidth]{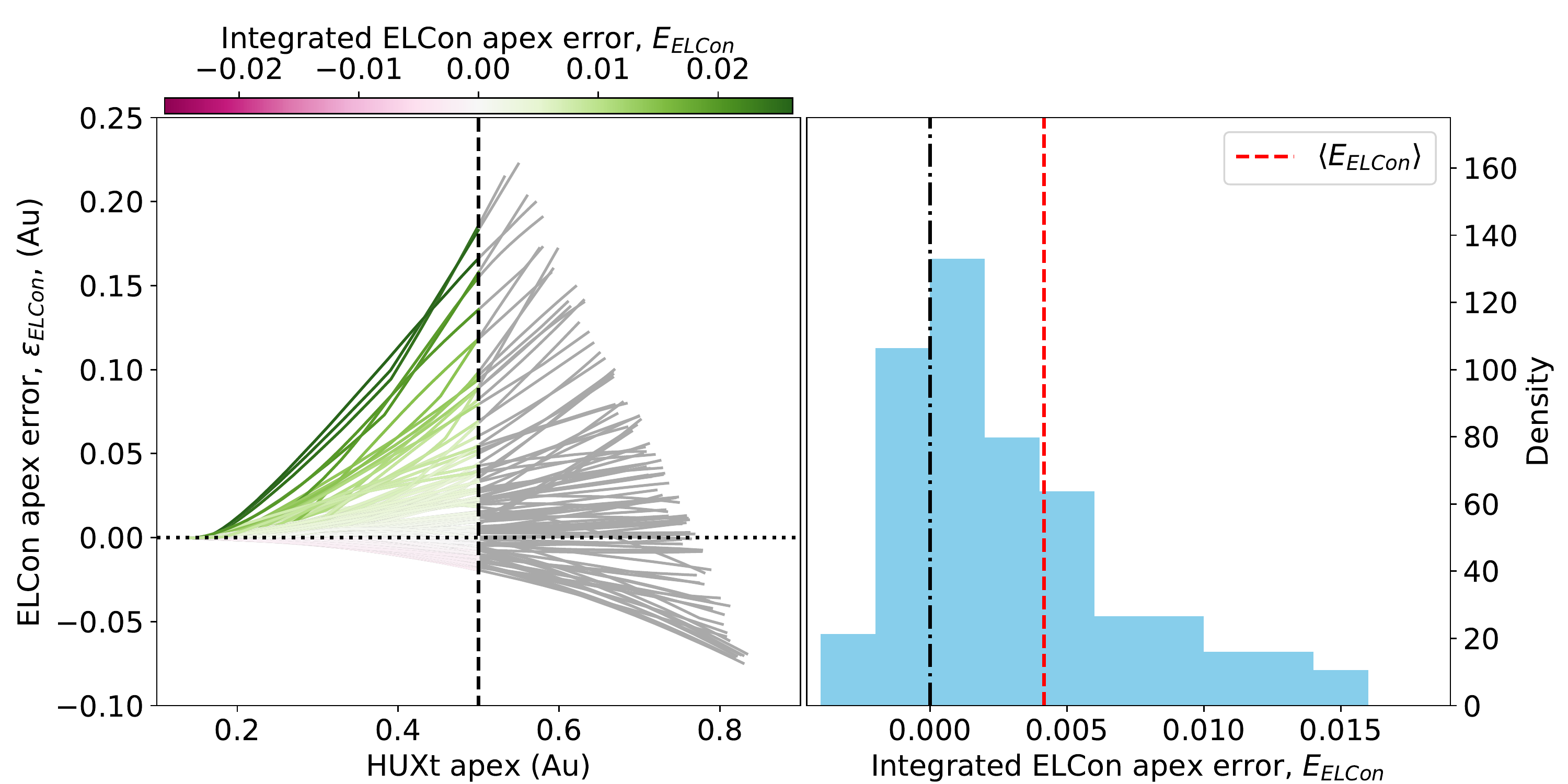}
\caption{(Left) This plot shows the error in the geometrically modelled CME apex distance as a function of the true CME apex distance, for the ELCon model with an observer at $270^{\circ}$ HEE longitude, for each of the 100 ambient solar wind solutions. To enable comparison between these error profiles across each of the geometric models and observer locations, they are integrated out to a maximum distance of 0.5~AU, as marked by the black dashed line. The color of each solid line indicates the sign and magnitude of the integrated error and the black dotted line marks the zero error line. (Right) A histogram of the integrated errors computed in the left-hand panel. The red dashed line shows the mean of the integrated error values, while the black dot-dash line marks the zero integrated error line.}
\label{fig:f06}
\end{figure}

Figure \ref{fig:f06} shows an example of computing $e_{ELCon}$ and $E_{ELCon}$ for the average CME scenario with an observer at $270^{\circ}$. The grey lines in the left hand panel show the $e_{ELCon}$ error values as a function of the true apex distance, while the vertical dotted line shows the $0.5$~AU integration limit. Below this limit the lines are colored according to the $E_{ELCon}$ value of each error series. The right hand panel shows a histogram of the 100 $E_{ELCon}$ values for this combination of CME scenario and observer location. The red dashed line shows the mean of the $E_{ELCon}$ values, $\langle E_{ELCon}\rangle$. Positive $e_{gm}$ values indicate that a geometric model predicts a CME apex distance larger than the true value and so positive $E_{gm}$ values represent a net bias of a geometric model to predicting CME apex distances that are too large. In the example of Figure \ref{fig:f06} it is clear that most of the simulations have positive $E_{ELCon}$ values, with a distribution that is skewed to the right, having a sharp fall in density below the modal bin, and a heavier tail above the modal bin. Consequently, the mean value, $\langle E_{ELCon}\rangle$, is positive, implying that on average, for an observer at $270^{\circ}$ and the average CME scenario, the ELCon model predicts CME-apex distances that are too large. 

With these summary statistics, we can now compare the geometric modelling results across different observer locations and CME scenarios. 

\begin{figure}
\includegraphics[width=\textwidth]{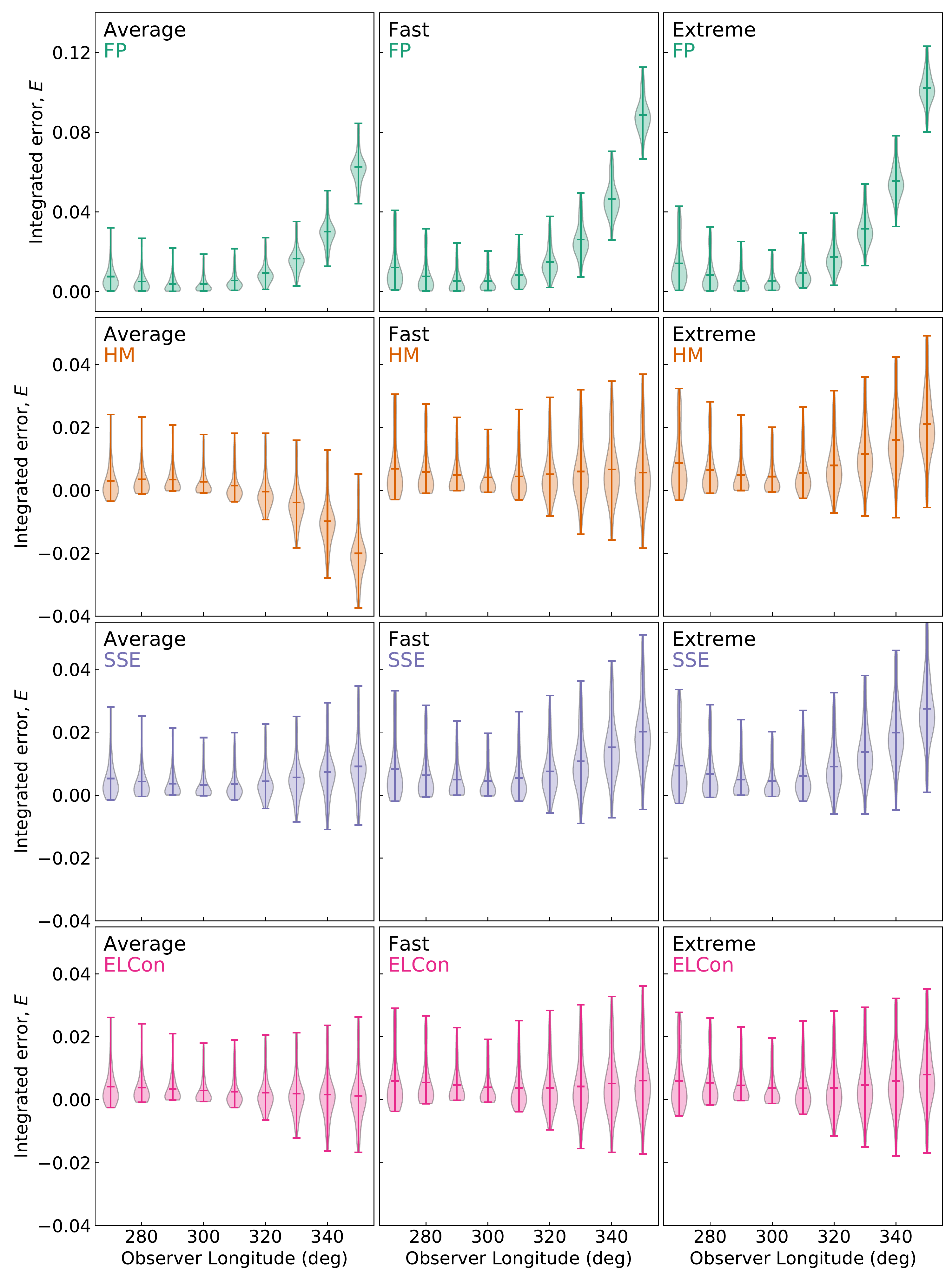}
\caption{These panels of violin plots show how the distributions of the integrated error, $E$, vary as a function of observer longitude, for each of the geometric models. The three columns correspond to the average, fast, and extreme CME scenarios. The four rows correspond to the FP, HM, SSE, and ElCon geometric models. Within each distribution, the horizontal bar marks the mean value. Please note that the bottom 3 rows all share the same y-axis limits, but that to sensibly fit the data to the panel, the top row required different limits.}
\label{fig:f07}
\end{figure}

\add[LB]{Figure} \ref{fig:f07} \add[LB]{presents the distributions of the integrated error, $E$, as a function of observer longitude, for each geometric model and CME scenario. These distributions are presented as violin plots, in which the shape of the distribution is estimated through kernel-density estimation and shown by the shaded region. The mean value is marked with the horizontal line near the center of each distribution. This shows that the shape and location of the distribution of $E$ changes significantly as a function of observer longitude. Generally, the distributions span mostly positive values of $E$, with the exception of the HM model for the average CME scenario, and observer longitudes $>320^{\circ}$. This suggests that, in general, geometric models are biased towards predicting CME apex distances that are larger than the true value. For each combination of geometric model and CME scenario, the spread of these distributions appears to have a minimum at observer longitudes around $300^{\circ}$. Furthermore, the spread appears to increase more rapidly as the observer longitude increases towards $350^{\circ}$ than it does as it decreases towards $270^{\circ}$. From this we conclude that, relative to observer longitudes in the range $270^{\circ}$-$350^{\circ}$, the L5 region likely offers the smallest distributions of uncertainties from geometric modelling}

\begin{figure}
\includegraphics[width=\textwidth]{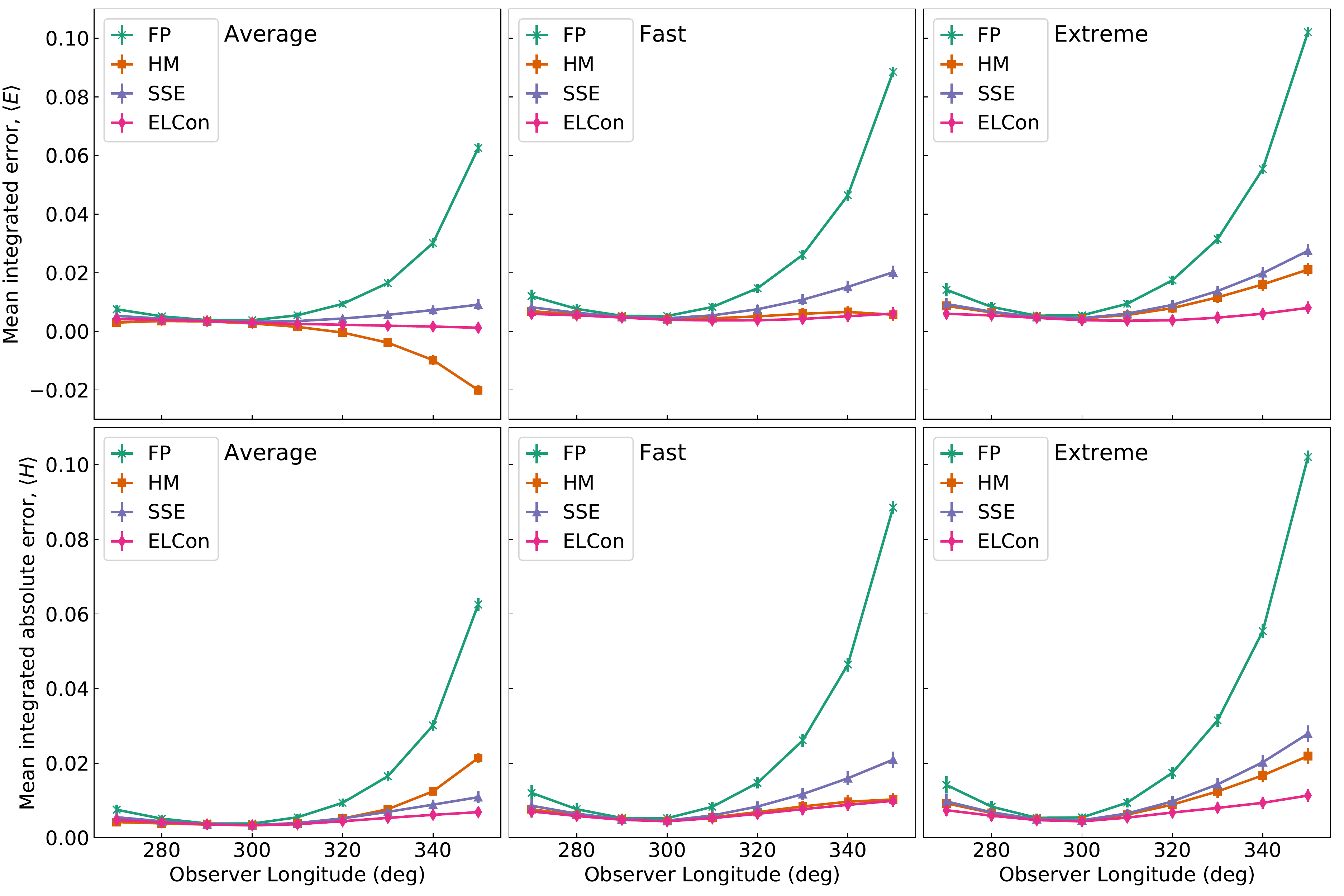}
\caption{These panels show how the mean integrated error, $\langle E\rangle$, and the mean integrated absolute error, $\langle H\rangle$, vary as a function of observer longitude, for each of the geometric models. The top row shows the $\langle E\rangle$ variation, and the bottom row shows the $\langle H\rangle$ variation. The three columns correspond to the average, fast, and extreme CME scenarios.}
\label{fig:f08}
\end{figure}

Figure \ref{fig:f08} \change[LB]{presents}{compares} the variation in $\langle E_{gm}\rangle$ and $\langle H_{gm}\rangle$ for the each CME scenario and range of observer longitudes. The columns correspond to the average, fast, and extreme CME scenario, while the top and bottom rows show the $\langle E_{gm}\rangle$ and $\langle H_{gm}\rangle$ variations, respectively. In each panel the colored lines with different marker styles show the variations $\langle E_{gm}\rangle$ and $\langle H_{gm}\rangle$ for each geometric model. Error bars are included that mark the range of 2 standard errors of the mean. Across all experiments, $\langle E_{gm}\rangle$ is positive, except for $\langle E_{HM}\rangle$ at observer longitudes $>320^{\circ}$. Therefore, in general, geometric models are biased towards predicting CME apex distances that are larger than the true value. We think this is because it is more likely that the flanks of CMEs interact with fast wind streams, which tends to advance the flanks relative to the CME apex. This results in an over-estimate of the CME apex radial distance when using the observed elongation of the flank to locate a simple geometric shape representing the CME front.

Of the geometric models, ELCon has the smallest magnitude $\langle E_{gm}\rangle$ and $\langle H_{gm}\rangle$ values across all observer longitudes. This suggests that, on average, the ELCon model is most successful at reconstructing the CME apex distance. The $\langle E_{gm}\rangle$ and $\langle H_{gm}\rangle$ values have a clear minimum value at an observer longitudes around $300^{\circ}$. Furthermore, in the region of this minima, the $\langle E_{gm}\rangle$ and $\langle H_{gm}\rangle$ values are all very similar, suggesting that, on average, all models perform similarly well in this observing configuration. Looking further away from the minima in $\langle E_{gm}\rangle$ and $\langle H_{gm}\rangle$, we see that the average errors grow faster with increasing observer longitude (decreasing Observer-Sun-CME angle). In particular, the errors on the FP model increase much more rapidly than with the other geometric models. In relation to the expected deployment of an operational space weather monitor to the L5 region, we note that this analysis provides some evidence that the errors of geometrically modelling Earth-directed CMEs are minimised for an observer in the L5 region.

Finally, we observe that the profiles of $\langle E_{gm}\rangle$ and $\langle H_{gm}\rangle$ are comparable across the three CME scenarios, particularly for the SSE and ELCon models. This suggests that the CME speed and width do not, on average, have a significant impact on how well the geometric models are able to reconstruct the CME apex distance. However, we also note that, in practice, observational issues such as feature tracking, motion blur, and feature distance from the Thomson plateau, might mean CME speed and width do have a significant impact on the performance of geometric modelling.

So far, this analysis of our simulations has demonstrated that solar wind structure is an important source of uncertainty in the estimation of a CMEs kinematics with geometric models. We now aim to provide a simple quantification of how the amount of solar wind structure relates to the error of the geometrically modelled CME kinematics. 

There is no standard metric for the level of structure in the solar wind, and it is possible to imagine the construction of a wide range of measures that focus on different aspects of solar wind structure. Here we choose to use a simple metric, which is the standard deviation of the solar wind speed on the HUXt inner boundary, over the longitude domain of the CME, at the timestep before CME initiation, $\sigma_{V_{b}}$. This quantifies the initial level of CME-front distortion expected from the ambient wind. Defining $V_{b}(\lambda_{i},t_{l})$ as the HUXt inner boundary speed values over longitude steps $\lambda_{i}$ and at the time step prior to CME launch, $t_{l}$, then we compute $\sigma_{V_{b}}$ as 

\begin{equation}
    \sigma _{V_{b}} = \left (\sum_{i \in i_{cme}} \left (V_{b}(\lambda_{i},t_{l}) - \langle V_{b}(\lambda_{i},t_{l}) \rangle_{i_{cme}} \right )^{2} / N_{i_{cme}}\right )^{\frac{1}{2}},
\end{equation}

where $i_{cme}$ are the longitude indices within the CME domain, $N_{i_{cme}}$ is the number of longitude indices spanned by the CME, and $\langle V_{b}(\lambda_{i},t_{l}) \rangle_{i_{cme}}$ is the mean of $V_{b}(\lambda_{i},t_{l})$ over the longitude indices $i_{cme}$.

\begin{figure}
\includegraphics[width=\textwidth]{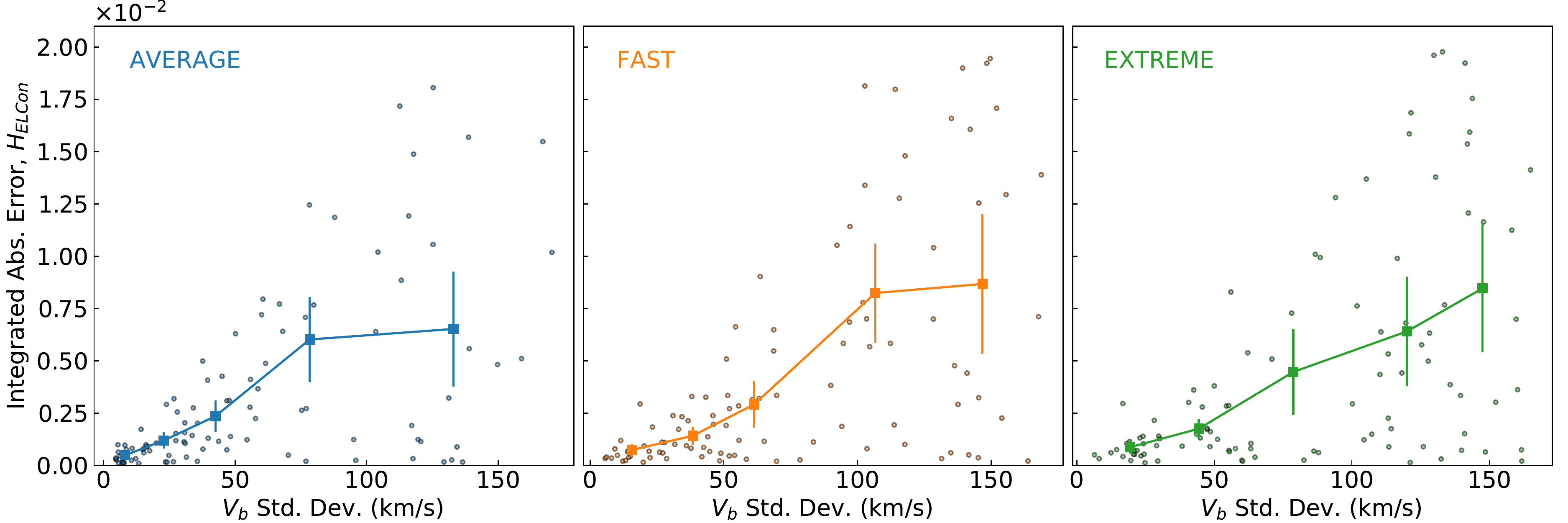}
\caption{The relationship between solar wind structure, quantified as the standard deviation of the HUXt solar wind speed inner boundary initial condition ($\sigma_{V_{b}}$), and the integrated absolute error in the ELCon kinematics estimates for an observer at $300^{\circ}$. The three columns correspond to the average, fast, and extreme CME scenarios. In each plot, the points mark the values for each of the 100 different background solar wind HUXt solutions. These data are then split by the quintiles of $\sigma_{V_{b}}$, and the squares and error bars show the mean values and two standard errors of the mean in each quintile.}
\label{fig:f09}
\end{figure}

Figure \ref{fig:f09} plots the relationship between $\sigma_{V_{b}}$ and $H_{ELCon}$ for each of the three CME scenarios, and for an observer at a longitude of $300^{\circ}$. Here we focus on the ELCon geometry because so far it has been shown to have the best performance, and on the observer longitude of $300^{\circ}$ because of its relevance for a future L5 mission. In each panel, the points correspond to each of the 100 different HUXt inner boundary conditions. These data are further split into bins corresponding to the quintiles of $\sigma_{V_{b}}$, and the mean values in each bin are computed. These data are marked by the squares, with the error bars corresponding to two standard errors of the mean.

For each CME scenario, it is clear that larger values $H_{ELCon}$ are more probable with increasing $\sigma_{V_{b}}$. However, the distributions show clear heteroskedasticity, with the variance in $H_{ELCon}$ being clearly conditional on $\sigma_{V_{b}}$. In particular, the lower limit of the $H_{ELCon}$ distribution does not seem to depend on $\sigma_{V_{b}}$; low errors on the geometrically modelled kinematics can be attained for all levels of solar wind structure. However, the upper limit of $H_{ELCon}$ grows rapidly with $\sigma_{V_{b}}$, and so the probability of there being large errors on the geometrically modelled kinematics increases quickly with increasing solar wind structure.

\subsection{ELEvoHI CME arrival times}

So far our analysis has only considered how well geometric models are able to reconstruct the kinematic profiles of CMEs with observations of the CME flank elongation. However, in isolation these techniques cannot provide a forecast of CME arrival, as they do not provide a means extrapolating forward in time the CME position. As introduced in section \ref{sec:ELEvoHI}, ELEvoHI couples the DBM to the ELCon geometry, and can therefore provide CME arrival time estimates. Here we assess the performance of the ELEvoHI CME arrival time estimates at Earth, using the suite of HUXt results for each CME scenario and background wind solution.

\begin{figure}
\includegraphics[width=\textwidth]{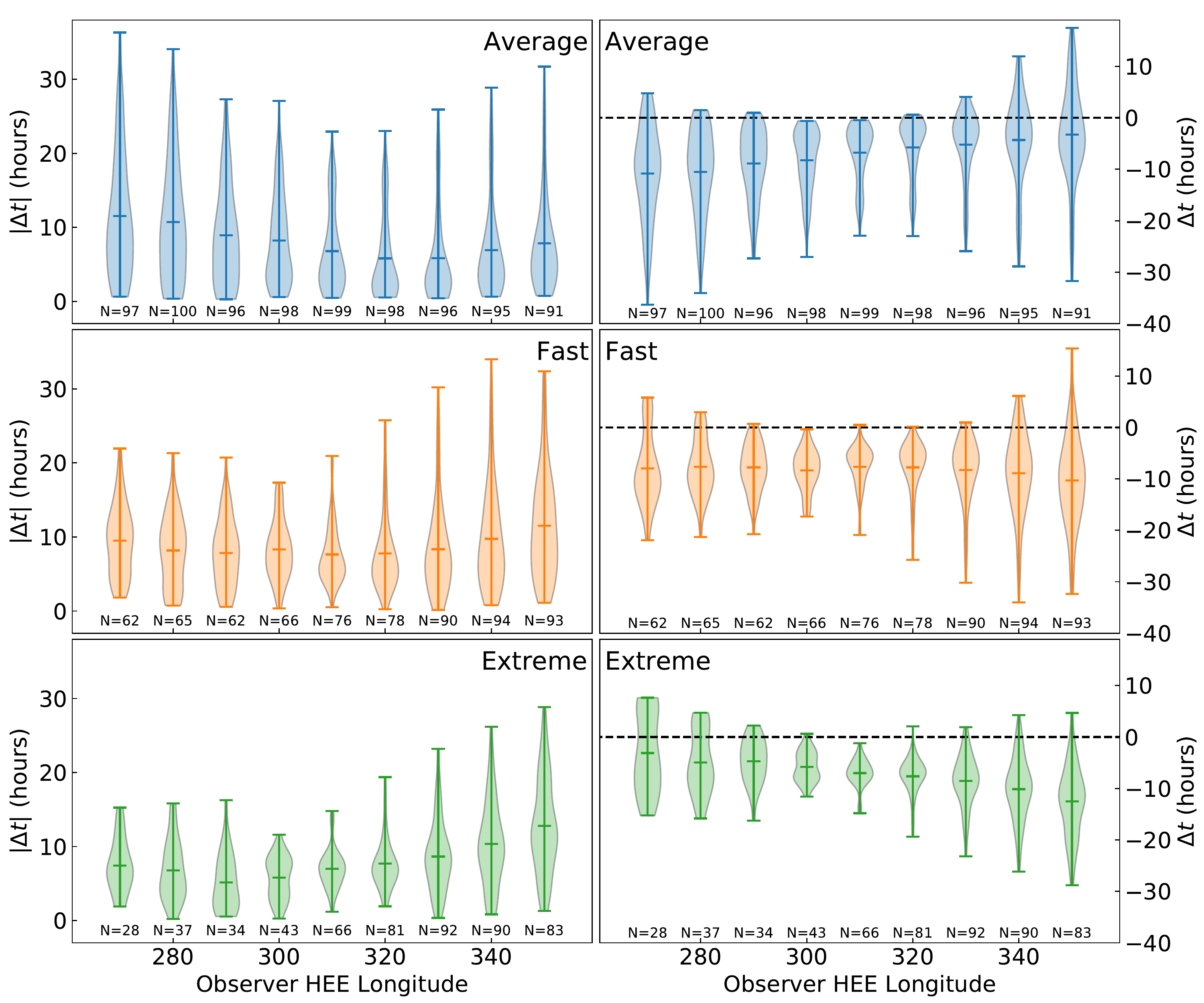}
\caption{These violin plots show the distributions of arrival time errors from the ELEvoHI predictions as a function of observer longitude and CME scenario. The left column shows the distributions of the absolute arrival time errors ($|\Delta t|$), while the right column shows the distributions of the arrival time errors ($\Delta t$). The rows correspond to the average, fast, and extreme CME scenarios. For a particular longitude, the shape of the error distribution is given by the violin. The horizontal line within the violin marks the mean of the distribution. The black dashed horizontal line marks the zero value of $\Delta t$. Positive $\Delta t$ correspond to a late arrival time.}
\label{fig:f10}
\end{figure}

To compare the ensemble ELEvoHI results with the simulated CME evolution, we compute the ensemble mean ELEvoHI arrival time, $\langle t \rangle$, and we characterise the performance of each ELEvoHI ensemble by computing the error of the ensemble mean arrival time as

\begin{equation}
\Delta t = \langle t \rangle - t_{huxt}
\end{equation}

such that positive $\Delta t$ corresponds to the ELEvoHI result arriving later at Earth than the true CME. Figure \ref{fig:f10} shows the distributions of the arrival time error, $\Delta t$, and absolute arrival time error, $|\Delta t|$, as a function of observer longitude, for each CME scenario. \add[LB]{Similar to the formatting of Figure } \ref{fig:f07}, these distributions are \add[LB]{also} presented as violin plots, in which the shape of the distribution is estimated through kernel-density estimation and shown by the shaded region. The mean value is marked with the horizontal line near the center of each distribution. The $\Delta t$ and $|\Delta t|$ distributions span ranges of approximately $-40$ to $10$~hours and $0$ to $40$~hours, respectively, depending on the CME scenario and observer location. For each CME scenario and observer longitude, most of the $\Delta t$ values are negative, indicating that ELEvoHI has a strong bias towards predicting early arrivals. We expect that this is related to the bias in the ELCon geometry to predicting CME apex distances that are too large. These distributions are also a function of observer longitude, and we observe that distribution spread is smallest at around $300^{\circ}$ longitude, corresponding to the L5 region. The spread of the distribution increases as we move away from the L5 region towards both larger and smaller $\phi$ angles, but particularly so for smaller $\phi$ angles. Related to this, most of the late arrivals correspond to the average and fast CME scenarios at $\phi < 20^{\circ}$, or for the extreme scenario at $\phi > 80^{\circ}$.

Considering now $|\Delta t|$, the spread of the distributions is clearly a function of observer longitude and CME scenario. The average, fast, and extreme CME scenarios have minima in the spread of $|\Delta t|$ at observer longitudes of $320^{\circ}$, $300^{\circ}$ and $300^{\circ}$ respectively. For each scenario, there is an asymmetry in how the spread of $|\Delta t|$ grows as we move away from the L5 region. For the average scenario, the spread increases more quickly as the $\phi$ angle increases towards $90^{\circ}$. However, for the fast and extreme scenario, the spread in $|\Delta t|$ increases more quickly as the $\phi$ angle decreases towards $10^{\circ}$.

There is a significant caveat in the analysis and interpretation of these ELEvoHI CME arrival statistics. Not all of the synthetic time-elongation profiles could be successfully fit with ELEvoHI in its current configuration. The number of samples is listed below each of the distributions in Figure \ref{fig:f10}. For the average scenario, ELEvoHI returned arrival times for almost all the HUXt runs and observer locations, with a minimum of 91/100. However, for the fast and extreme scenarios this fraction drops significantly, and appears to be a function of observer longitude. For the fast scenario, the minimum number of samples is 62/100 at an observer longitude of $270^{\circ}$, rising steadily to a maximum of 93/100 at an observer longitude of $350^{\circ}$. The same pattern is seen in the extreme scenario, but with minima and maxima of 28/100 and 83/100, respectively. So it is important to understand that our conclusions on the performance of ELEvoHI are conditional on ELEvoHI returning an arrival time, and that this probability might be a strong function of CME scenario. We have analysed the runs that could and could not be fit with ELEvoHI to try and understand what features of a CMEs kinematics could cause it to fail to converge on a solution. However, so far there are no clear systematic differences between the HUXt runs that did and did not result in a successful ELEvoHI fit. Further analysis will be necessary to understand this behaviour.

As discussed in \citeA{verbeke_benchmarking_2019}, it is typical to reduce the CME arrival time error distributions into some summary metrics that aid the comparison of different CME forecasting techniques. Following \citeA{verbeke_benchmarking_2019}, we compute the mean arrival time error ($\langle \Delta t\rangle$), mean absolute arrival time error ($\langle |\Delta t|\rangle$), root mean square error (RMSE, $RMSE_{\Delta t}$), and standard deviation ($\sigma_{\Delta t}$), as a function of CME scenario and observer longitude. Each metric provides different information on the shape and location of the arrival time error distribution. The mean error is a useful measure of the bias in the forecasts, of whether the forecasts typically predict an early or late arrival. The mean absolute error is a commonly used metric for assessing the skill of a forecast. The RMSE provides similar information to the mean absolute error, but gives more weight to larger errors and is therefore more sensitive to outliers. The standard deviation is used as a measure of the spread of the distributions.

\begin{figure}
\includegraphics[width=\textwidth]{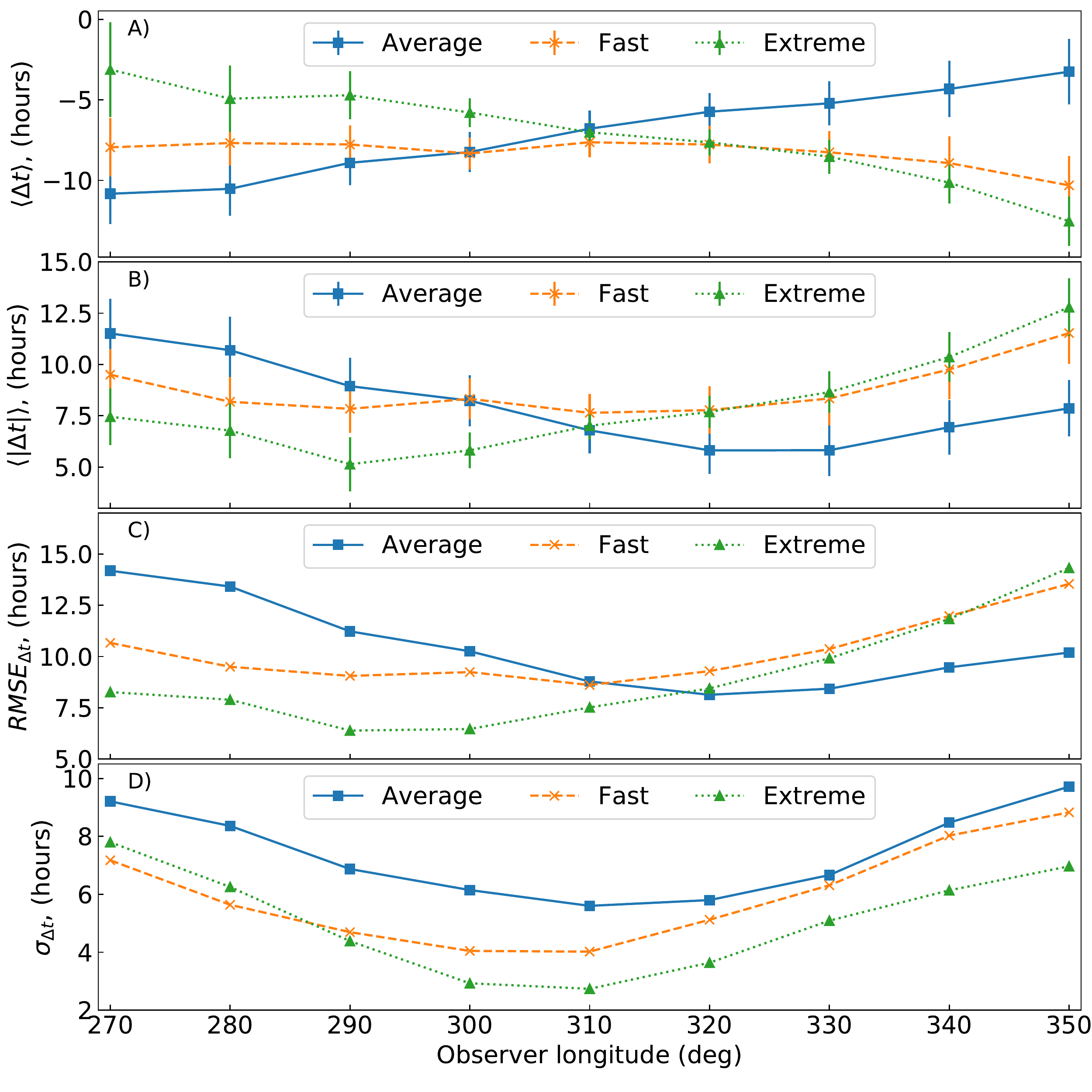}
\caption{Series of error metrics computed from the ElEvoHI arrival time error distributions, as a function of CME scenario and observer longitude. (A) The mean error. (B) The mean absolute error. (C) The root-mean-square error. (D). The standard deviation. The average, fast, and extreme series are shown with blue squares, orange stars, and green triangles. For the mean error, and mean absolute error, the uncertainty bars are 2 standard errors of the mean.}
\label{fig:f11}
\end{figure}

Figure \ref{fig:f11} presents these error metrics. Considering first $\langle \Delta t\rangle$, it is clear that there are trends in $\langle \Delta t\rangle$ with observer longitude, but that these trends depend on the CME scenario. For the average scenario, $\langle \Delta t\rangle$ is larger in magnitude at higher $\phi$ angles, and decreases in magnitude monotonically with decreasing $\phi$ angle. For the fast scenario, the variation in $\langle \Delta t\rangle$ with $\phi$ angle is flat with no clear trend. While the extreme scenario shows the opposite trend to the average scenario, with the average error being smallest at large $\phi$ angles and monotonically increasing in magnitude with decreasing $\phi$ angle.

The $\langle |\Delta t|\rangle$ and $RMSE_{\Delta t}$ series do not show the same trends with observer longitude as do the $\langle \Delta t\rangle$ series. In fact, the variations of $\langle |\Delta t|\rangle$ and $RMSE_{\Delta t}$ are not monotonic, and show minima occurring at intermediate values of observer longitude. The location of these minima appears to vary systematically with the CME scenario, occurring at $320^{\circ}$, $310^{\circ}$, and $290^{\circ}$ for the average, fast, and extreme CME scenarios respectively. We interpret this as evidence that the optimal $\phi$ angle for an observer increases as the speed and width of the CME increase. We suspect it is the CME width that is critical in driving this behaviour, but cannot confirm this with our chosen CME scenarios. Future work could consider a wider range of CME scenarios to decouple the effects of CME speed and width.

The standard deviation series also show minima occurring at intermediate values of observer longitude. However, in this instance the location of the minima seems to be equal for each CME scenario, at an observer longitude of $310^{\circ}$.

At a longitude of $300^{\circ}$, in the L5 region, the mean absolute arrival time error for the average, fast, and extreme CME scenarios which could be fitted with ELEvoHI are $8.2\pm 1.2~h$, $8.3\pm 1.0~h$, and $5.8\pm 0.9~h$, respectively. The average across all scenarios, weighting by the sample size in each scenario, is $7.8~h$. Similarly, the standard deviations are $6.2~h$, $4.1~h$, and $2.9~h$, for the average, fast, and extreme CME scenarios, respectively. We again note that for the fast and extreme scenarios, caution should be taken in interpreting these values, as only $66\%$ and $43\%$ of the fast and extreme simulations could be fit with ELEvoHI; therefore these average errors are conditional on events that could be successfully fitted. Nonetheless, with this limitation in mind, we note that in the L5 region, these average values are similar across all three CME scenarios, and comparable to the empirically established uncertainties from the ELEvoHI validation studies of around $6-7~h$ \cite{Rollett2016,amerstorfer_evaluation_2021,hinterreiter_why_2021}. As our study does not include any observational uncertainty and as we see no reason why the model and observational uncertainties would compensate each other, we expect that the true uncertainty should be larger than calculated here. In this context, we note that although the empirically established ELEvoHI uncertainties are comparable to those computed in our study, they are in fact systematically smaller than ours. We suggest that there are two obvious factors that could influence this. Firstly, the modest sample sizes of the validation studies could mean they have not properly sampled the uncertainty distribution yet. This is not a criticism of these studies, as the relative infrequency of CMEs through the STEREO mission limits the available sample of well observed Earth-directed CMEs. Secondly, it could be that the representation of cone CMEs in HUXt is overestimating the impact of solar wind structure on CME evolution. This is quite plausible, as the CMEs are purely hydrodynamic velocity perturbations with no magnetic structure and we expect that a CMEs magnetic structure would typically serve to inhibit the rate of CME deformation by structured solar wind. Although we also note that in many circumstances CMEs are not expected to behave as coherent magnetic structures \cite{Owens2017c}. It would be beneficial to compare the kinematics of cone CMEs in HUXt with magnetised CMEs in a 3D MHD model, to assess whether such a consideration is important. However, at present we lack the resources to repeat our experiment completely with a 3D MHD model and magnetised CMEs.

\begin{figure}
\includegraphics[width=\textwidth]{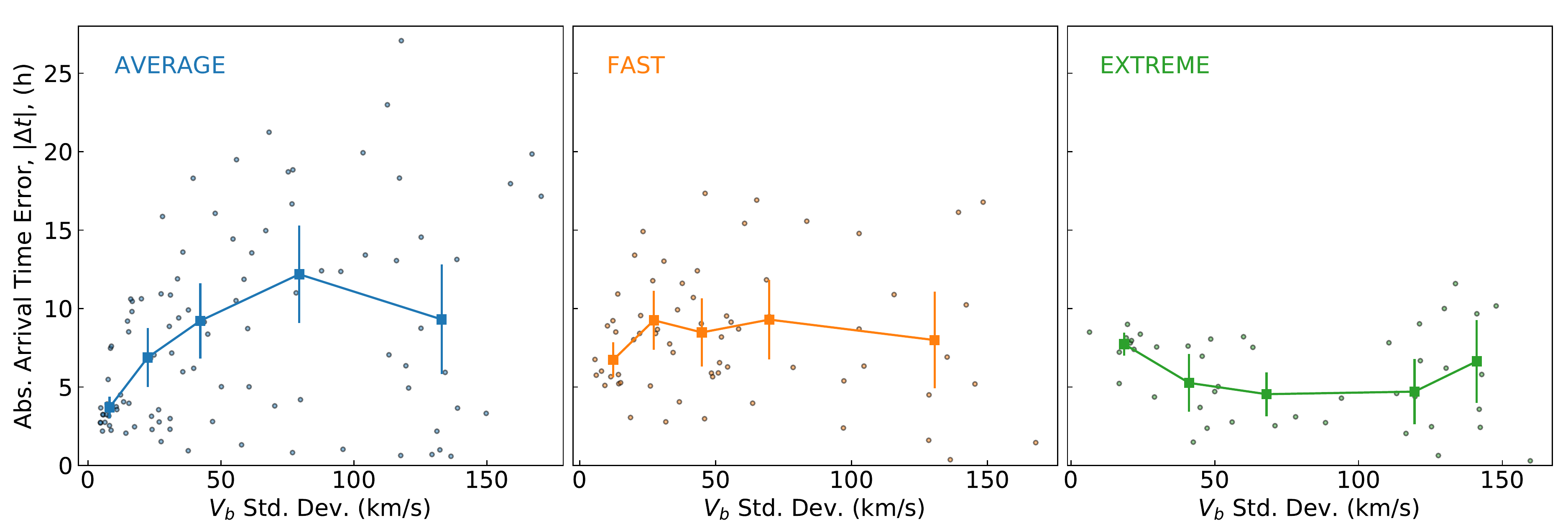}
\caption{These plots show the relationship between the standard deviation of the HUXt solar wind speed inner boundary initial condition and the absolute error on the ELEvoHI arrival time, for an observer at $300^{\circ}$. The three columns correspond to the average, fast, and extreme CME scenarios. In each plot, the points mark the values for each of the 100 different background solar wind HUXt solutions. These data are then split by the quintiles of $\sigma_{V_{b}}$, and the squares and error bars show the mean values and two standard errors of the mean in each quintile.}
\label{fig:f12}
\end{figure}

Although we have demonstrated that the distribution of ELEvoHI arrival time errors does depend on both the structure of the solar wind and the observer location, we have not quantified how the level of structure in the solar wind affects the magnitude of the ELEvoHI arrival time errors. Here, we again use $\sigma_{V_{b}}$ as a measure of the level of solar wind structure, and compare these values against the $|\Delta t|$ corresponding to the L5 observer at a longitude of $300^{\circ}$. Figure \ref{fig:f12} shows these data for each of the three CME scenarios, with the points showing the values corresponding to each of the available background solar wind solutions. In the same way as with Figure \ref{fig:f09}, we split these data into bins based on the quintiles of the $\sigma_{V_{b}}$ distribution, and compute the means of the values in each bin. These mean values are shown by the squares, with the errors bars corresponding to two standard errors of the mean. For the average CME scenario we observe that the absolute arrival time error does tend to increase with increasing $\sigma_{V_{b}}$ but, again, these data are heteroskedastic, with the variance in $|\Delta t|$ growing with $\sigma_{V_{b}}$. The lower limit of the $|\Delta t|$ distribution does not appear to be a function of $\sigma_{V_{b}}$, suggesting that low ELEvoHI arrival-time errors are possible for all levels of solar-wind structure with the average CME scenario. However, these data also show that larger ELEvoHI arrival-time errors become increasingly probable with increasing $\sigma_{V_{b}}$.

Interestingly, we do not observe similar patterns in the $\sigma_{V_{b}}$ and $|\Delta t|$ relationship for the fast and extreme CME scenarios. Within the limits of these samples, the distribution of $|\Delta t|$ appears to be approximately uniform with $\sigma_{V_{b}}$ which suggests there is no clear relationship between solar wind structure and the ELEvoHI arrival-time errors for these scenarios. This is somewhat surprising, particularly in the context of Figure \ref{fig:f09}, which demonstrated that there was a relationship between $\sigma_{V_{b}}$ and the ELCon kinematics errors for each CME scenario, and which are a key component of the ELEvoHI modelling results. The cause of this is unclear to us, but we suggest one plausible explanation. We note that ELEvoHI was only successfully fit to a subset of the HUXt runs, with arrival-time values for only $66\%$ and $43\%$ of the runs in the fast and extreme CME scenarios. Therefore, it is plausible that this result is due to a selection bias, based on the possibility that the probability of ELEvoHI failing to return an arrival- time estimate could be larger when the error on the ELCon kinematics increases. Such an effect would naturally exclude samples with large ELCon errors that would presumably also correspond to large arrival-time errors.

\section{Conclusions}
\label{sec:conclusions}

Geometric modelling of CME kinematics is a widely used tool within space-weather research and is being actively developed as a means of providing CME arrival-time forecasts from heliospheric imagery. Yet the necessary assumptions of such models, coupled with the challenges of tracking CMEs through heliospheric imagery, means that there are significant uncertainties in geometrically modelled CME kinematics estimates. One source of uncertainty that has been thought to be particularly significant is that geometric models typically neglect the impact of solar wind structure on the evolution of the CME, as solar wind structure is understood to significantly affect CME propagation \cite{case_ambient_2008}. Our study aimed to use simulations to quantify the scale of uncertainty introduced into CME geometric modelling by solar wind structure.

To do this, we developed three cone CME scenarios representing an average, a fast, and an extreme CME, \change[LB]{and we}{that are directed along Earth's latitude and longitude, with no inclination to the ecliptic plane. We then} used the HUXt solar wind model to simulate the interaction of these cone CME scenarios with 100 different time-dependant background solar wind environments. For each simulation, synthetic time-elongation profiles of the CME flanks were generated from virtual observers at a range of heliospheric locations relative to Earth and these are assumed to be representative of those that are derived from heliospheric imager data,. With these synthetic data, we computed the geometrically modelled kinematics of each simulated CME scenario, and compared these with the ``true'' CME kinematics. This analysis revealed several key findings:

\begin{itemize}
    \item The Elliptical Conversion (ELCon) geometry typically performs better than the Fixed-Phi (FP), Harmonic Mean (HM) and Self-Similar-Expansion (SSE) geometries, having the lowest overall mean errors in reconstructing the CME kinematics.
    
    \item In the low heliosphere, it is often the case that the geometric models return essentially degenerate estimates of the CME apex distance. In this context it is unclear whether it is advantageous to use simpler geometries with less free parameters (e.g. FP and HM), or more complex geometries with additional free parameters (e.g. SSE and ELCon).
    
    \item For most combinations of geometric model and observer location, the geometric models are biased towards predicting CME apex distances that are larger than the true value. 
    
    \item For an Earth-directed CME, the lowest mean geometric modelling error is returned for an observer at around a HEE longitudes of $300^{\circ}$, which corresponds to the L5 Lagrange region.
    
    \item The magnitude of the CME speed and width do not appear to significantly affect the mean error of the \change[LB]{geometrically modelled CME kinematics}{modelled CME kinematics for any of the tested geometric models}. 
    
    \item The mean and variance of the geometric modelling errors increase with increasing solar wind structure.
\end{itemize}

Regarding the general bias of geometric models towards predicting CME apex distances that are larger than the true value, we suggest that this is because solar wind structure tends to advance the flanks of the CME relative to the apex, flattening the CME front. This type of behaviour was observed in the case studies presented in Figures \ref{fig:f03} and \ref{fig:f04}. This is also consistent with the result that the ELCon geometry returned the lowest overall mean error; the flatter CME front used by the ELCon geometry serves to reduce the over-estimation of the CME apex distance.

We also analysed the impact of solar wind structure on the ELEvoHI arrival time estimates. This revealed that:

\begin{itemize}
    \item Arrival time errors vary between 15 hours late and 35 hours early, with a strong bias towards early arrival time estimates. 
    
    \item The arrival-time-error distributions vary as a function of both observer location and CME scenario. For each CME scenario, there is a minimum in the mean absolute arrival-time error, $|\Delta t|$ as a function of observer longitude. This minima was located at $320^{\circ}$, $310^{\circ}$, and $290^{\circ}$ longitude for the average, fast, and extreme CME scenario, respectively.
    
    \item For an observer in the L5 region, the mean arrival time error is around $8~h$, but also depends on CME scenario, being $8.2\pm1.2~h$, $8.3\pm1.0~h$, and $5.8\pm0.9~h$ for the average, fast, and extreme scenarios, respectively.
    
    \item There is some evidence that the ELEvoHI arrival-time errors increase with increasing solar-wind structure for the average CME scenarios. However there is no clear evidence that the arrival-time errors depend on the level of solar-wind structure for the fast and extreme CME scenarios.
    
\end{itemize}

An important caveat in the interpretation of the ELEvoHI arrival time estimates is that we could only obtain valid ELEvoHI estimates for a subset of $66\%$ and $43\%$ of the fast and extreme CME profiles. Future work should consider what aspects of the simulated CME evolution were challenging for ELEvoHI to represent. \add[LB]{Furthermore, as a CME's speed is a critical factor in determining it's impacts on Earth, our future work will extend this study to also analyse the uncertainty on the CME arrival speeds predicted by the geometric models and ElEvoHI.}

In this study we have focused on the impact of solar-wind structure on the uncertainty of geometrically modelled CME kinematics, absent of observational errors. It would be useful future work to also consider the impact of observational errors, which would allow the computation of a more complete uncertainty budget. We expect the total uncertainty should be larger than estimated from this simulation study, as we see no reason why the observational and solar-wind structure driven uncertainties should compensate each other.

Our simulation results provide some evidence that solar-wind structure is a significant source of uncertainty in geometrically modelled kinematics. This supports the results of \citeA{hinterreiter_why_2021}, who concluded that solar wind structure was the main reason for differences in ELEvoHI arrival time predictions for the same CMEs when fitted with either the STEREO-A or STEREO-B HI data.

Our results also show that it could be possible to estimate the likely level of uncertainty in the geometrically modelled kinematics from a quantification of the level of solar-wind structure in the heliosphere. Such an approach could help assess the plausibility of the geometrically modelled kinematics, which could be useful in both research and forecast settings.

\section{Data}
The HUXt model was obtained from \url{https://github.com/University-of-Reading-Space-Science/HUXt}.

The analysis code and data supporting this study is available from  \url{https://github.com/University-of-Reading-Space-Science/GeoModelUncertainty} and an archived citable version is provided by Zenodo at \url{https://doi.org/10.5281/zenodo.5552110}

\acknowledgments
This work was in part supported by the SWIGS NERC Directed Highlight Topic Grant NE/P016928/1/. T.A., J. H., and C.M. thank the Austrian Science Fund (FWF): P31265-N27, P31659-N27, P31521-N27.
This research has made use of SunPy v1.1.3 \cite{sunpysoftware}, an open-source and free community-developed solar data analysis Python package \cite{sunpypaper}, and also of Astropy, a community-developed core Python package for Astronomy \cite{astropy:2013, astropy:2018}. Graphics were made using Matplotlib \cite{matplotlib}.

%% ------------------------------------------------------------------------ %%
%% References and Citations

\end{document}